\documentstyle[aps,graphicx,preprint,tighten]{revtex}

\draft

\newcommand{\be}{\begin{equation}}
\newcommand{\ee}{\end{equation}}
\newcommand{\bea}{\begin{eqnarray}}
\newcommand{\eea}{\end{eqnarray}}
\newcommand{\nn}{\nonumber}
\newcommand{\qc}{\quad ,}
\newcommand{\qp}{\quad .}
\newcommand{\rmi}{{i}}
\newcommand{\rme}{{\rm e}}

\newcommand{\cly}{{\cal Y}}

\newcommand{\eh}{{\frac{1}{2}}}
\newcommand{\drh}{{\frac{3}{2}}}
\newcommand{\fh}{{\frac{5}{2}}}

\newcommand{\sss}{$N(1535)$ }
\newcommand{\ddd}{{$N(1520)$} }

\newcommand{\bsl}[1]{#1 \!\!\! /}
\newcommand{\eq}[1]{Eq.~(\ref{#1})}
\newcommand{\eqs}[1]{Eqs.~(\ref{#1})}
\newcommand{\msc}[1]{Sec.~\ref{#1}}
\newcommand{\fig}[1]{Fig.~\ref{#1}}
\newcommand{\figs}[1]{Figs.~\ref{#1}}
\newcommand{\grad}{\vec{\nabla}}

\def\Journal#1#2#3#4{{#1} {\bf #2} (#4) #3}
\def\NP{{\em Nucl. Phys.} }

\def\PL{{\em Phys. Lett.} }
\def\PR{{\em Phys. Rev.} }
\def\PRL{{\em Phys. Rev. Lett.} }

\begin{document}
%\begin{titlepage}
\begin{center}

{\Large \bf

Coherent Photoproduction of
Eta-Mesons on Spin-Zero Nuclei in a Relativistic, Non-local Model
\footnote{Work supported by BMBF and GSI Darmstadt.}
\footnote{This work forms part of the dissertation of W. Peters.}
}
\bigskip
\bigskip

{W. Peters\footnote{Wolfram.Peters@theo.physik.uni-giessen.de}, 
H. Lenske and U. Mosel}

\bigskip
\bigskip
{ \it
Institut f\"ur Theoretische Physik, Universit\"at Giessen\\
D-35392 Giessen, Germany\\ }

\end{center}

\begin{abstract}
  The coherent photoproduction of $\eta$-mesons on spin-zero nuclei is
  studied in a relativistic, non-local model, which we have previously
  applied to the coherent photoproduction of pions.  We find that
  different off-shell extrapolations of the elementary production
  operator lead to large effects in the cross section.  We also show
  that the almost complete suppression of the $N(1535)$ seen in
  earlier studies on this reaction is a result of the local or
  factorization approximation used in these works. Non-local effects
  can lead to a considerable contribution from this resonance. The
  relative size of the $N(1535)$ contribution depends on the structure
  of the nucleus under consideration. We give an estimate for the
  contribution of the $N(1520)$ resonance and discuss the effect of an
  $\eta$-nucleus optical potential.  \pacs{PACS numbers: 25.20.Lj,
    24.10.Jv}
\end{abstract}

\section{Introduction}

Photonuclear reactions offer a unique possibility to test our
understanding of hadronic interactions in vacuum as well as in the
nuclear environment. For elementary processes like the photoproduction
of pions and $\eta$-mesons there are sophisticated microscopic models
available. These are based on an effective Lagrangian approach and
describe the experimental data accurately (see for example
\cite{feuster,mukop} and references therein).  These models provide us
with single-particle production operators, needed to treat
photonuclear reactions on nuclei in the impulse approximation. Among
the numerous possible reactions the coherent photoproduction, in which
the nucleus remains in the ground state after the production process,
plays a special role: Since the initial and the final state of the
nucleus are the same and well understood, the uncertainties with
respect to the nuclear structure are less important than in other
reactions.

The coherent photoproduction of $\eta$-mesons on spin-zero nuclei has
experienced renewed interest recently. After work based on a
multipole parameterization of the elementary amplitude
\cite{trya,bennhold,tiator}, new studies starting from effective
Lagrangian models for the elementary process appeared recently
\cite{fix,pieka}.  One reason for this renewed theoretical effort,
apart from the progress made with respect to our understanding of the
elementary production, is that in all these works the contribution of
the $N(1535)$, which dominates the elementary cross section, is found
to be strongly suppressed when coherent production on the nucleus is
considered. Thus the other contributions, resulting from the Born
terms, the \ddd resonance and the omega exchange, were predicted to be
clearly visible in the coherent photoproduction on nuclei.

In all these works the so called local or
factorization approximation is used. In this approximation, the
momentum of the incoming nucleon is fixed to a certain value, so that
the evaluation of the nuclear transition matrix elements is
simplified. In the present study we treat the coherent
photoproduction of $\eta$-mesons in a relativistic non-local model,
i.e. we take the dependence of the production
operator on the momentum of the incoming nucleon into account. This
model has successfully been applied to the coherent photoproduction of
pions on nuclei \cite{pions}.

In the following section we will first give the details of the
elementary model. We then describe how the resulting production
operator is evaluated in the case of coherent production.  In
\msc{results} the different contributions to the coherent cross
section as well as the properties of the non-local contributions
are discussed in detail. The effect of simple
$\eta$-nucleus optical potentials is shown. We finally summarize our
findings in \msc{sum}.

\section{The model}
\subsection{The elementary operator}
\label{theon}
\begin{figure}[!ht]
  \centerline{ \includegraphics[width=10cm]{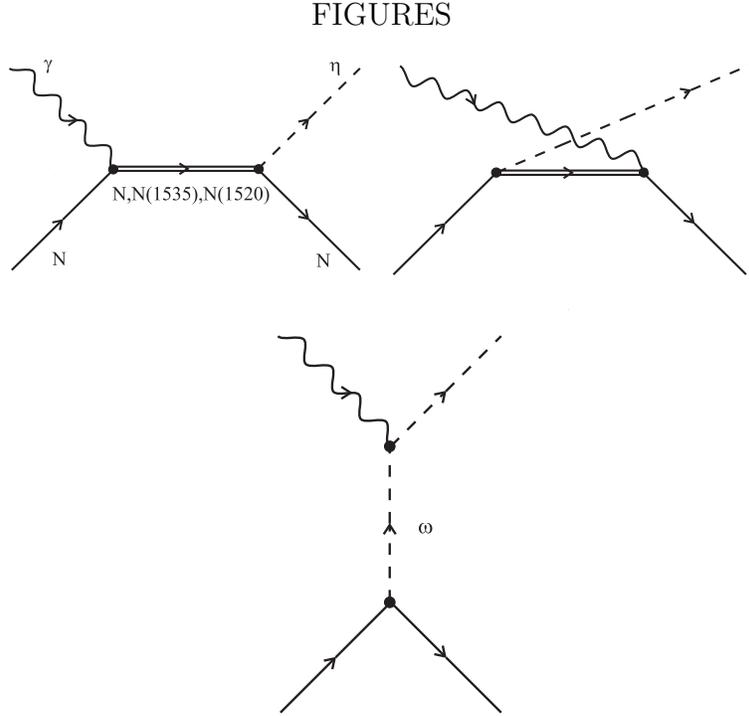} } 
  \caption{Feynman diagrams contributing to the photoproduction of 
$\eta$-mesons on a free nucleon.}
\label{diagrams} 
\end{figure}

The Feynman diagrams contributing to the photoproduction of $\eta$
mesons on a nucleon are shown in \fig{diagrams}. These graphs are
constructed from the following interaction Lagrangians:
\bea
\label{lint} 
{\cal L}_{\eta NN} &=& - i\:{g_{_{\eta NN}}} \: 
                       \overline{\psi}_N \: \gamma_5
                       \: \psi_N\: \eta
                      \nn\\
                      \nn\\
{\cal L}_{\gamma NN} &=& - {e} \: 
                       \overline{\psi}_N \: {\textstyle \frac{1}{2}}(1+\tau_3) 
                       \gamma_{\mu} \: 
                      \: \psi_N \: A ^\mu \nn\\
                      & & - {\textstyle \frac{1}{2}} \:{\rm e} \:
                      \frac{\kappa_p}{ 2 m_{_N}} \:
                      \overline{\psi}_N \: {\textstyle \frac{1}{2}}(1+\tau_3)\:
                      \sigma_{\mu \nu} \:\psi_N \:F^{\mu \nu }\nn\\
                      & & - {\textstyle \frac{1}{2}} \:{\rm e} \:
                      \frac{\kappa_n}{ 2 m_{_N}} \:
                      \overline{\psi}_N \: {\textstyle \frac{1}{2}}(1-\tau_3)\:
                      \sigma_{\mu \nu} \:\psi_N \:F^{\mu \nu }\nn\\
                      \nn\\
{\cal L}_{\eta NS_{11}} &=& -i \:{g_{_{\eta NS_{11}}}} \: 
                       \overline{\psi}_N 
                       \: \psi_{S_{11}}\: \eta + h.c.
                      \nn\\
                      \nn\\
{\cal L}_{\gamma NS_{11}} &=& \frac{eg_{\gamma N S_{11}}}{4m_N}
                      \:
                      \overline{\psi}_{S_{11}} \: \gamma_5\:
                      \sigma_{\mu \nu} \:\psi_N \:F^{\nu \mu }+ h.c.\nn\\
                      \nn\\
                      \nn\\
{\cal L}_{\eta ND_{13}} &=&  \:\frac{g_{_{\eta ND_{13}}}}{m_\eta} \: 
                       \overline{\psi}_{D_{13}}^\mu \: \gamma_5
                       \: \psi_N\: \partial_\mu  \vec{\eta}+ h.c.
                      \nn\\
                      \nn\\
{\cal L}_{\gamma ND_{13}}^{(1)} &=&  \:
                      i\frac{e g_{_{\gamma ND_{13}}}^{(1)}}{2m_N} \: 
                       \overline{\psi}_{D_{13}}^\mu \:\gamma^\nu
                       \: \psi_N\: F_{\mu\nu}+ h.c.
                      \nn\\
                      \nn\\
{\cal L}_{\gamma ND_{13}}^{(2)} &=&  \:
                      \frac{e g_{_{\gamma ND_{13}}}^{(2)}}{4m_N^2} \: 
                       \overline{\psi}_{D_{13}}^\mu \:\partial^\nu
                       \: \psi_N\: F_{\mu\nu}+ h.c.
                      \nn\\
                      \nn\\
{\cal L}_{\omega \eta \gamma} &=&  
                       \frac{e g_{_{\omega \eta \gamma}}}{4m_\eta} \: 
                       \varepsilon_{\mu\nu\rho\sigma}  
                       \:(V^{\mu\nu} \: F^{\rho\sigma})\eta\nn\\
                      \nn\\
{\cal L}_{\omega NN} &=& -  g_{\omega NN}^v \: 
                       \overline{\psi} \: 
                       \gamma_{\mu} \: 
                      \: \psi \: \omega ^\mu \nn\\
                      & & - {\textstyle \frac{1}{2}} \:  
                      \frac{g_{\omega NN}^t}{ 2 m_{_N}} \:
                      \overline{\psi} \: 
                      \sigma_{\mu \nu} \:\psi \:V^{\mu \nu }
\qc
\eea
with $ F_{\mu\nu} = \partial_\mu A_\nu - \partial_\nu A_\mu$ and $
V_{\mu\nu} = \partial_\mu \omega_\nu - \partial_\nu \omega_\mu$, where
$A_\mu(\omega_\mu)$ denotes the photon (omega) field.

We take the resonance parameters, as well as the $\omega$ and nucleon
parameters, from \cite{fix}, where the data for the elementary
photoproduction cross section \cite{bernd} are well reproduced.  The
corresponding coupling constants are given in Tab.~\ref{cc}.  In
addition, a form factor is introduced at the $\omega NN$-vertex in
\cite{fix}:

\bea
\label{omff}
F(t) &=& \frac{\Lambda^2 - m_\omega^2}{\Lambda^2 -t}
\qc
\eea
with $\Lambda^2$ = 1.2 GeV$^2$, which is also used here.

\section{Photoproduction on the nucleus}
\label{theonucl} 

\subsection{The nuclear wave function}
\label{nucwav}

The wave functions needed in the following for the bound nucleons are
taken from a relativistic mean-field calculation using scalar and
time-like vector potentials $V_s$ and $V_v$, respectively:

\bea
 \label{nuceom}
(\bsl p\: - \: m \: -\: V_v \gamma_o \: -\: V_s) \psi_\alpha \:= \: 0
\qp
\eea
For the potentials $V^v $ and $V^s$ we assume a Woods-Saxon shape:

\bea
\label{nucpot}
V(r) \: = \: V_i^o\: \left( 1\: +\: 
      \rme^{\frac{(r - r_iA^{1/3})}{a_i }}\right)^{-1}
\quad ; i = v,s
\qp
\eea
The parameters for these potentials are given in Table \ref{nucpotp}.
They were determined such that the separation energies, the root mean
square radius of the charge density and the charge form factors of
$^{12} {\rm C}$ and $^{40} {\rm Ca}$ are well reproduced up to a
momentum transfer of 3 fm$^{-1}$ \cite{pions}.  
{ As discussed in
\cite{pions}, nuclear correlations, which are neglected in a
mean-field approach, can become important at a momentum transfer
larger than 3 fm$^{-1}$ \cite{who}.}

\subsection{The amplitude for the coherent photoproduction on the nucleus}
\label{theocoh}

The elementary model described in \msc{theon} together with the single
particle nuclear wave functions introduced in the previous section are
now used to calculate the cross section for the coherent
photoproduction of $\eta$ mesons on the nucleus. In the impulse
approximation (IA) this is done assuming that the production process
takes place on a single nucleon, neglecting many-body contributions.
Thus one has to evaluate the diagrams in \fig{diagrams} replacing the
wave functions of in- and outgoing nucleons by the bound state wave
functions.  The resulting amplitude $T$ is then related to the
differential cross section for the coherent photoproduction of $\eta$
mesons via \bea
\label{cs}
\frac{d\sigma}{d\Omega} \:=\: \left( \frac {M_A}{4 \pi W}\right)^2 \:
                     \frac{q_{cm} }{  k_{cm}}\:\frac{1}{2}\:
                     \sum\limits_\lambda\: \mid T^{(\lambda)} \mid^2
\qc
\eea
where $M_A$ is the mass of the nucleus, $k_{cm}$ and $q_{cm}$ denote
the three-momentum of the photon and the $\eta$, respectively, in the
cm-frame. The total cm-energy is denoted by $W$ and $\lambda$ stands for the
photon polarization.

Since a bound nucleon is off-shell, the matrix element of the
production operator has to be evaluated for kinematical situations
different from those in the free case. In our model this is done by
calculating the corresponding matrix element with the production
operator taken directly from the Feynman diagrams.

We work in position space, since the bound state and scattering
wave functions are easily obtained in a position space representation.
As an example, the direct diagram involving the \sss resonance
corresponds in our approach to the following non-local expression:

\bea
\label{amplex}
T_{S_{11} }^{(\lambda)} &=&
\sum\limits_{\alpha \:occ.} \: \int d^3 x \: d^3 y \:
      \overline{ \psi}_\alpha(\vec{x}) \: {\phi_{\eta}^{(-)}}^*(\vec{x}) \:
      \Gamma_{\eta NS_{11}}\:G_{S_{11}} (E;\vec{x},\vec{y})\:
      \Gamma_{\gamma NS_{11}}^\mu\:
      \phi_\mu^{(\lambda)} (\vec{y})\: \psi_\alpha(\vec{y})
\nn \\
&=&
\sum\limits_{\alpha \:occ.} \: \int d^3 x \: d^3 y \:
      \overline{ \psi}_\alpha(\vec{x}) \: {\phi_{\eta}^{(-)}}^*(\vec{x}) \:
      \hat{T}_{S_{11}}^{(\lambda)}(E;\vec{x},\vec{y})\:
      \phi_\gamma (\vec{y})\: \psi_\alpha(\vec{y})
\qp
\eea
Here $\psi_\alpha$ is the wave function of the bound nucleon.
$\phi_\mu^{(\lambda)} = \varepsilon_\mu^{(\lambda)} \phi_\gamma$ is
the wave function of the photon, where $\phi_\gamma$ is
a plane wave. 
$\phi_{\eta}^{(-)}$ is the wave function of
the $\eta$ satisfying incoming boundary conditions \cite{joach}.
$\Gamma_{\eta NS_{11}}$ and $\Gamma_{\gamma NS_{11}}^\mu$ are the
vertices resulting from the coupling terms in \eq{lint} and
$G_{S_{11}}$ is the resonance propagator:

\bea
\label{s11prop}
G_{S_{11}} (p_o;\vec{x},\vec{y}) = \int \: \frac{d^3p}{(2\pi)^3} \:
{\rmi \rme^{\rmi \vec{p} ( \vec{x}-\vec{y})}} \frac{\bsl p + 
  m_{S_{11}}}{p^2 - m_{S_{11}}^2 + i m_{S_{11}} \Gamma(p^2)} \qp 
\eea 
The
energy dependence of $\Gamma(p^2)$ is taken to be: 
\bea
\Gamma(p^2)=\Gamma_o \left( b_\eta \frac{q_\eta}{q_\eta^o} + 
  b_\pi \frac{q_\pi}{q_\pi^o} + b_{\pi\pi} f(p^2)\right) 
\qp
\eea 
For the branching ratios we take $b_\eta$ = 0.5, $b_\pi$ = 0.4 and
$b_{\pi\pi}$ = 0.1 and the width is $\Gamma_o$=160 MeV.  $q_\eta$
($q_\pi$) stands for the three momentum of the $\eta$ (the pion) in
the rest frame of the resonance as a function of its invariant mass
$p^2$, $q_\eta^o$ ($q_\pi^o$) denote the respective momenta on the mass
shell of the resonance. { The function $f(p^2)$ fullfils 
$f(m_{S_{11}}^2)$=1 and contains the energy dependence of
the two-pion decay as given in \cite{carrasco}. }

The energy $E$ in \eq{amplex} is naturally determined by energy conservation:
\bea
E = E_\gamma + E_\alpha
\qc
\eea
where $E_\alpha$ is the total, relativistic energy of the bound nucleon.

The resonance propagator depends in position space on $\vec{x}$ and
$\vec{y}$ independently, so that the nucleon wave functions in
\eq{amplex} are evaluated at different positions.  This introduces a
non-locality in \eq{amplex}.  
This non-locality corresponds to a
process where a nucleon is taken out of the nucleus at position
$\vec{y}$, and subsequently put back at position $\vec{x}$.
We thus have to calculate a
six-dimensional integral; for the technical details we refer the
reader to Ref.~\cite{pions}.

For the following discussion it will be useful to rewrite
\eq{amplex}, assuming for the moment the $\eta$ to be 
represented by a plane wave:
\bea
\label{amplextr}
T_{S_{11} }^{(\lambda)} 
&=&
  \int d^3 x \: d^3 y \:Tr\left[
       \:
      \hat{T}_{S_{11} }^{(\lambda)}(E;\vec{x},\vec{y})\:
       \hat{\rho}_A(\vec{y},\vec{x}) \right]
    \: \rme^{\rmi \vec{k} \vec{y}} \rme^{-\rmi \vec{q} \vec{x}}
\qp
\eea
Here $\hat{\rho}_A(\vec{x},\vec{y})$ denotes the
nuclear density matrix in position space:
\bea
\label{rhoxy}
\hat{\rho}_A(\vec{x},\vec{y}) =
\sum\limits_{\alpha \:occ.} \: 
      \psi_\alpha(\vec{x}) \otimes
      \overline{ \psi}_\alpha(\vec{y})
\qc
\eea
which is a non-local quantity.  The symbol $\otimes$ denotes the
dyadic product of the two spinors.  From the local parts
$\hat{\rho}_A(\vec{x},\vec{x})$ of this density matrix, the usual
scalar and vector and tensor densities of the nuclear ground state are
obtained via:
\bea
\label{svt}
\rho_s(\vec{x})&=&Tr[\hat{\rho}_A(\vec{x},\vec{x})]
\quad {} \quad
\rho_v(\vec{x})=Tr[\gamma_o\:\hat{\rho}_A(\vec{x},\vec{x})] \nn\\
\rho_t(\vec{x})\hat{\vec{x}}_i&=&Tr[\sigma^{oi}\:\hat{\rho}_A(\vec{x},\vec{x})]
\qp
\eea
In \cite{fix} and \cite{pieka}, the production operator is treated
differently.  In both studies the production operator is projected
onto a minimal set of Lorentz-covariant operators using the free Dirac
equation \cite{mukop}.  In \cite{fix} its matrix elements are then
evaluated assuming that in the nucleus the relation between small and
large components of the nucleon spinors is the same as in free space.
The nuclear structure enters in \cite{fix} via empirical
parameterizations for the vector form factor of the nucleus.  In
\cite{pieka} a different set of Lorentz-covariant operators is used
and the nucleon wave functions are taken from a relativistic
mean-field calculation like in our model.  In this study the
production operator is treated relativistically and consists, after
the projection procedure, of a tensor, a pseudoscalar and a
pseudovector part.  In the local approach of \cite{pieka} it is found,
that only the tensor part yields a
non-vanishing contribution. Thus in \cite{pieka} the coherent process
only involves the tensor density of the nuclear ground state
(\eq{svt}). This will be discussed further in \msc{omega}.

Since both these ways of evaluating matrix elements of the production
operator involve the use of on-shell relations, they are equivalent
for the $\eta$ production on a free nucleon, in fact they represent
standard techniques used to calculate the elementary process.  In the
case of off-shell nucleons, however, they lead to different results.
Thus, despite of starting from rather similar elementary models, the
two works \cite{fix} and \cite{pieka} differ with respect to the
off-shell behavior from each other and from our approach.  We avoid
this ambiguity by taking the production operator directly from the
Feynman diagrams, without rewriting it in any way.  Thus, we use the
natural off-shell behavior of an effective field theory. It must of
course be kept in mind, that this also represents an extrapolation of
a microscopic model to a kinematical region where it has not been
tested against experiment.

The second important difference between \cite{fix} and \cite{pieka} on
one side and our approach on the other side is the treatment of the
dependence of the production operator on the momenta of the in- and
outgoing nucleons.  In \cite{fix,pieka} the local or factorization
approximation is used, which amounts to putting the momentum
$\vec{p}$ of the incoming nucleon equal to a fixed value $\vec{p}_o$.
In order to obtain this approximation,
\eq{amplex} must be
transformed to momentum space. Again we assume
the $\eta$ wave function to be a plane wave:
\bea
\label{amplexmom}
T_{S_{11} }^{(\lambda)} &=&
\sum\limits_{\alpha \:occ.}  
\int  \frac{d^3 p}{(2\pi)^3} \: 
      \overline{ \psi}_\alpha(\vec{p}+\vec{k}-\vec{q}) \:
      \Gamma_{\eta NS_{11}}\:G_{S_{11}} (E;\vec{p}+\vec{k})\:
      \Gamma_{\gamma NS_{11}}^{(\lambda)}\:
      \psi_\alpha(\vec{p}) \nn \\
 &=& 
\sum\limits_{\alpha \:occ.} \: 
\int  \frac{d^3 p}{(2\pi)^3} \: 
      \overline{ \psi}_\alpha(\vec{p}+\vec{k}-\vec{q}) 
      \:\hat{T}_{S_{11} }^{(\lambda)} (  E;\vec{p},\vec{k},\vec{q})\:
      \psi_\alpha(\vec{p})
\qc
\eea
where we have set $\Gamma_{\gamma NS_{11}}^{(\lambda)} = 
                   \Gamma_{\gamma NS_{11}}^\mu \varepsilon_\mu^{(\lambda)}$.
In this equation
$\vec{k}$($\vec{q}$) denotes the three-momentum of the photon (the $\eta$).
The local approximation now amounts to  neglecting the dependence of 
$\hat{T}_{S_{11} }(  E;\vec{p},\vec{k},\vec{q})$ on the nucleon
momentum $\vec{p}$. This is done by the replacement $\vec{p} \to
\vec{p}_o$, where $\vec{p}_o$ 
{ is independent of the integration variable $\vec{p}$, and is
often assumed to depend on the momentum transfer \cite{pions}.}
Thus one
puts:
\bea
\label{amplexloc}
T_{S_{11} }^{(\lambda)} 
 &\to& Tr \left[
         \hat{T}_{S_{11} }^{(\lambda)} ( E;\vec{p}_o,\vec{k},\vec{q}) \:\:
         \hat{\rho}_A(\vec{k}-\vec{q}) \right]
\qc
\eea
with
\bea
\label{rhoa}
\hat{\rho}_A(\vec{p}) &=&  
\int  \frac{d^3 p'}{(2\pi)^3} \: 
\sum\limits_{\alpha \:occ.} \: 
      \psi_\alpha(\vec{p}\:') \otimes
      \overline{ \psi}_\alpha(\vec{p}\:'+\vec{p}) \nn \\
 &=&
\int {d^3 x} \: \rme^{\rmi \vec{p} \vec{x}} \:
\sum\limits_{\alpha \:occ.} \: 
      \psi_\alpha(\vec{x}) \otimes
      \overline{ \psi}_\alpha(\vec{x})
=\int {d^3 x} \: \rme^{\rmi \vec{p} \vec{x}} \: \hat{\rho}_A(\vec{x},\vec{x})
\qc
\eea
with $\hat{\rho}_A(\vec{x},\vec{x})$ being the local part of the nuclear
density matrix from \eq{rhoxy}. The wave functions of in- and
outgoing nucleons are now evaluated at the same position.

We now read off from \eqs{amplextr}, (\ref{amplexloc}) and
(\ref{rhoa}) that the local approximation corresponds in position space to
the replacement:
\bea
\label{locpos}
\hat{T}_{S_{11} }^{(\lambda)}(E;\vec{x},\vec{y}) \to 
 \hat{T}_{S_{11} }^{(\lambda)} ( E;\vec{p}_o,\vec{k},\vec{q})
 \delta^3(\vec{x}-\vec{y})
\qc
\eea
which is the approximation used in Refs.~\cite{fix,pieka}. 
A similar formula for the local approximation has been given in 
\cite{johan}.
This
approximation is widely used in DWIA calculations,
 in a non-relativistic framework its validity
has been studied in the photoproduction of charged pions in
\cite{li,tiator2}.

In this context, the $\omega$-exchange plays a special role. The
contribution of the corresponding graph is in our approach given by:
\bea
\label{amplompos}
T_{\omega}^{(\lambda)} &=& \sum\limits_{\alpha \:occ.}  \int d^3 x \:
d^3 y \: \overline{ \psi}_\alpha(\vec{x}) \Gamma_{\omega NN}^\mu
\psi_\alpha(\vec{x}) \:G_{\mu\nu}^\omega(E;\vec{x},\vec{y})
\:\Gamma_{\omega \eta \gamma}^{\nu\sigma}
{\phi_{\eta}^{(-)}}^*(\vec{y}) \phi_\sigma^{(\lambda)} (\vec{y}) \nn\\
&=&  \int d^3 x \: d^3 y \:
Tr\left[\hat{\rho}_A(\vec{x},\vec{x}) \Gamma_{\omega NN}^\mu \right]
\:G_{\mu\nu}^\omega(E;\vec{x},\vec{y})
\:\Gamma_{\omega \eta \gamma}^{\nu\sigma}
{\phi_{\eta}^{(-)}}^*(\vec{y}) \phi_\sigma^{(\lambda)} (\vec{y})
\qc
\eea 
with $E=E_\gamma - E_\eta$.  The important difference between
\eq{amplompos} and Eqs.~(\ref{amplex}) and (\ref{amplextr}) is that in
the case of the $\omega$-graph, only the local part of the
density matrix $\hat{\rho}_A$ appears.
In momentum space \eq{amplompos} reads:
\bea
\label{amplom}
T_{\omega}^{(\lambda)} &=& \sum\limits_{\alpha \:occ.}  \int \frac{d^3
  p}{(2\pi)^3} \: \overline{ \psi}_\alpha(\vec{p}+\vec{k}-\vec{q}) \:
\Gamma_{\omega NN}^\mu\: \psi_\alpha(\vec{p})\:
G_{\mu\nu}^\omega(E;\vec{k}-\vec{q})\: \Gamma_{\omega \eta
  \gamma}^{\nu{(\lambda)}}
\nn\\
&=& \sum\limits_{\alpha \:occ.}  \int \frac{d^3 p}{(2\pi)^3} \:
\overline{ \psi}_\alpha(\vec{p}+\vec{k}-\vec{q}) \:
\hat{T}_{\omega}^{(\lambda)}(E;\vec{k},\vec{q}) \psi_\alpha(\vec{p})\:
\qp 
\eea 
where we have set $\Gamma_{\omega \eta \gamma}^{\nu{(\lambda)}}=
\Gamma_{\omega \eta
  \gamma}^{\nu\sigma}\varepsilon_\sigma^{(\lambda)}$.  Due to the
appearance of only the local part of the nuclear density matrix
$\hat{\rho}_A(\vec{x},\vec{x})$ in \eq{amplompos},
$\hat{T}_{\omega}^{(\lambda)}$ does not depend on the nucleon momentum,
so that the local approximation has no effect on this
amplitude. Consequently, for the omega contribution the different
off-shell behavior of the production operator is the main difference
between \cite{fix}, \cite{pieka} and our model.

\subsection{The eta-nucleus optical potential}
\label{etaoptsc}

It is well known that the production of mesons like pions and
$\eta$-mesons on the nucleus is strongly affected by the final state
interaction between the outgoing meson and the nucleus.  While this is
neglected in the plane wave impulse approximation (PWIA), it is
commonly taken into account in the frame work of the distorted wave
impulse approximation (DWIA), by introducing an optical potential, and
using scattering wave functions instead of plane waves in \eq{amplex}.
The pion-nucleus interaction is theoretically and experimentally
sufficiently well known, while for the $\eta$ meson there are no data
available at all, and we have to resort to theory. 

{ 
  
  We will show results obtained with the following optical potentials:
  The optical potential used in \cite{pieka} is based on the results
  of \cite{bennhold}. There the $\eta$-nucleon scattering amplitude is
  obtained using a multipole parameterization, which is then used to
  construct an $\eta$-nucleus optical potential via a simple
  $t\rho$-approximation.
  
  In \cite{carrasco} the inclusive $\eta$-photoproduction is studied using
  an $\eta$ self-energy that is based on the model developed in
  \cite{chiang}.  The corresponding optical potential goes beyond the
  $t\rho$-approximation by using an in-medium self-energy for the
  intermediate $N(1535)$, thus including terms of arbitrary order in
  the density.
  
  We finally constructed a third optical potential using the results
  of \cite{effe}. Within the models used in this study it was found,
  that in order to reproduce the data for the inclusive
  photoproduction of $\eta$-mesons, an energy-independent in-medium
  $\eta N$ cross section is needed. For the inelastic $\eta N$ cross
  section a value of 30 mb is given.  This can be converted into an
  optical potential via a $t\rho$-ansatz, assuming the real part to be
  zero.  We find \mbox{$-2\omega_\eta\:V_\eta=i \:3 \:fm^2 \:p_\eta\:\rho$}, where
  $\omega_\eta$ is the energy of the $\eta$ and $p_\eta$ its
  three-momentum.
  
  In the following the results of the different optical potentials
  will be labeled DWIA I, II and III, respectively.}  The coherent
photoproduction treated here might help to obtain more information
about the $\eta$-nucleus interaction.

\section{Results}
\label{results}
In this section the results of our calculations for the coherent
photoproduction of $\eta$ mesons on $^{12}{\rm C}$ and $^{40}{\rm Ca}$ are
discussed in detail. 
We will also give results for $^4{\rm He}$, despite of the limited
applicability of our model for this nucleus.
Since there are different aspects to be mentioned
for each graph, we will first discuss the differential cross
sections for each graph separately, and in the end give results for a
complete calculation, with and without an $\eta$-nucleus optical
potential.

\subsection{The omega graph and the Born terms}
\label{omega}

In the previous section it has been shown that the present study and
\cite{fix,pieka} differ the least in the case of the $\omega$-graph,
we therefore start our discussion with this term.  Due to the
local nature of this graph, the off-shell extrapolation of the
production operator is the only difference, the effect of which
can be seen in comparison to the results in \cite{fix,pieka}.

The
$\omega$ couples to the nucleon via a vector and a tensor coupling
(\eq{lint}).    Due to the relative size of the
coupling constants and the additional momentum dependence in the case
of the tensor coupling, the vector coupling strongly dominates.
For a pure vector coupling the Dirac structure of the corresponding
production operator is given by a simple vector term
(cf.~\eq{amplom}):
\bea
\label{omgammu}
\hat{T}_\omega^{(\lambda)} = a_\mu \gamma^\mu
\eea
with:
\bea
a_\mu = - i g_{\omega NN}^v 
G_{\mu\nu}^\omega(E;\vec{k}-\vec{q})\: \Gamma_{\omega \eta
  \gamma}^{\nu{(\lambda)}}
\qp
\eea
From this it is clear, that the $\omega$ couples directly to the
vector density of the nucleus.
The omega amplitude then fulfills:
\bea
\label{omt2}
\sum\limits_{\lambda} \mid T_\omega^{(\lambda)}\mid^2 = 
g^2 \frac{k^2 q^2 \sin^2\theta}{(t-m_\omega^2)^2} 
\mid F_v(Q)\mid^2
\qc
\eea
where $g$ is a factor containing the coupling constants and the
$\omega NN$ form factor. In this equation $k$ and $q$ stand for the
three-momentum of the photon and the $\eta$-meson, respectively,
$Q=\mid \vec{k}-\vec{q}\mid$ is the transfered momentum and $F_v(Q)$
is the Fourier transform of the vector density, i.e.~the vector
form factor:
\bea
\label{rhofou}
F_v(Q) = 4 \pi\:\int r^2 dr\: j_o(Qr) \rho_v(r) 
\qc 
\eea 
where $j_o$ is the spherical Bessel function of order 0 and $\rho_v$
is the vector density of the nuclear ground state.  Thus in our
approach the vector coupling of the $\omega$ leads to a cross section
that is directly proportional to the vector form factor.

This is in line with \cite{fix}, where the differential cross section
is also taken to be proportional to the vector form factor, but it is
in contrast to \cite{pieka}.
There the production operator in \eq{omgammu} is replaced by:
\bea
\label{ompieka}
\hat{T}_\omega^{(\lambda)} \to
F_T^{\alpha \beta} \sigma_{\alpha\beta} + F_A^{\alpha} \gamma_5\gamma_\alpha
\qp
\eea
The explicit expressions for $F_T^{\alpha \beta}$ and $F_A^{\alpha }$
are given in \cite{pieka}. Instead of the vector term in \eq{omgammu}
a tensor and a pseudovector term appear.  For the other graphs there
is an additional pseudoscalar term, which vanishes in the case of a
pure $\omega NN$ vector coupling \cite{mukop}.  Due to the local
nature of the $\omega$-term, only the tensor term in (\ref{ompieka})
can contribute to the coherent photoproduction, leading to the tensor
form factor in the cross section in \cite{pieka}.  To obtain this
tensor term from the original vector structure in \eq{omgammu}, one
has to use the free Dirac equation \cite{mukop}.  Thus a production
operator of the form (\ref{ompieka}) is equivalent to the original
vector term in \eq{omgammu} for the photoproduction on a free nucleon,
but shows a different behavior if the nucleon is off-shell.

\begin{figure}[!ht]
  \centerline{ \includegraphics[width=13cm]{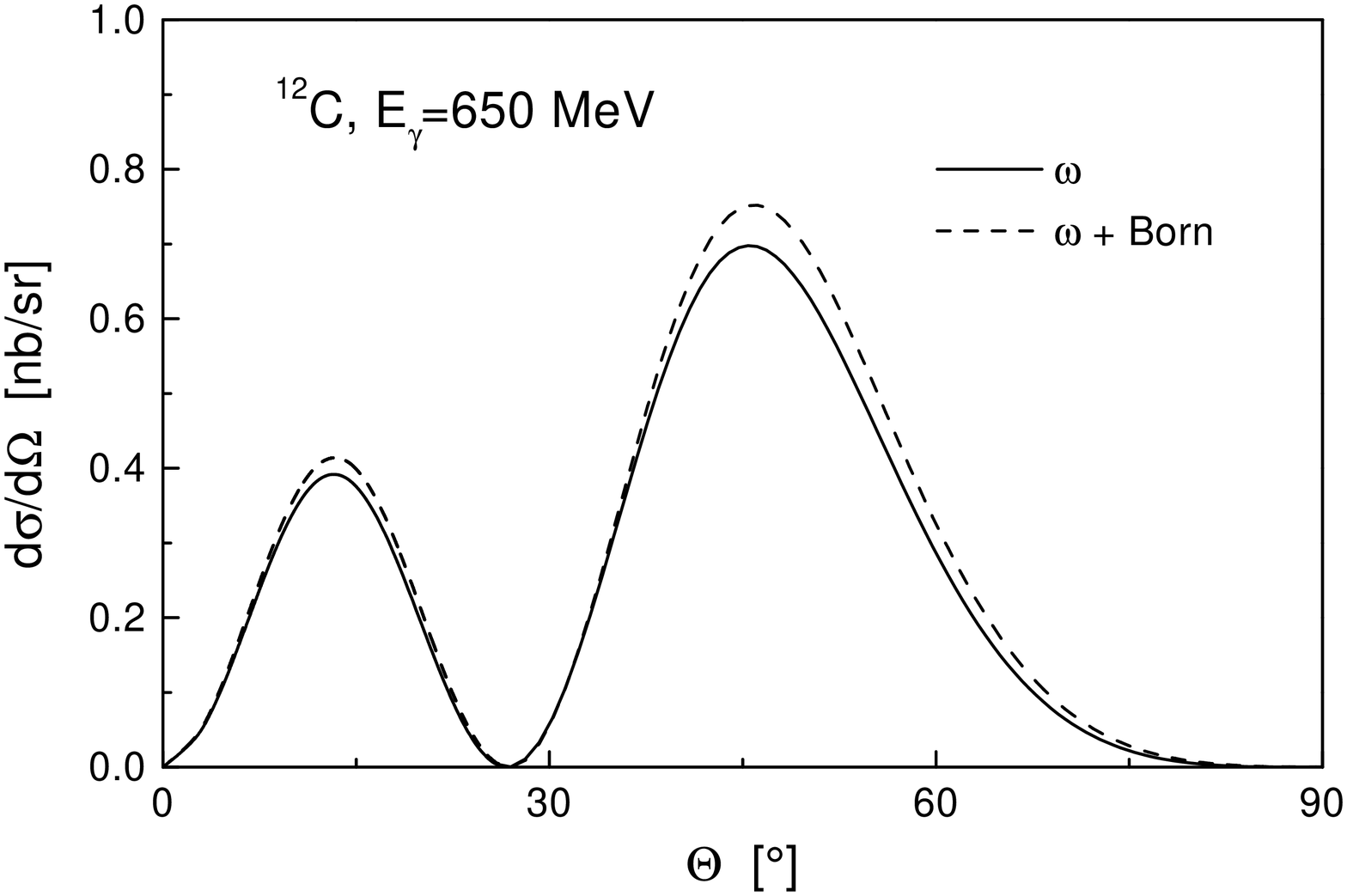} }
  \caption{Differential cross section for the coherent photoproduction
    of $\eta$-mesons on $^{12}{\rm C}$ for a photon laboratory energy of 650
    MeV resulting from the omega graph and the nucleon Born terms. 
}
\label{dsccom} 
\end{figure}

\begin{figure}[!ht]
  \centerline{ \includegraphics[width=13cm]{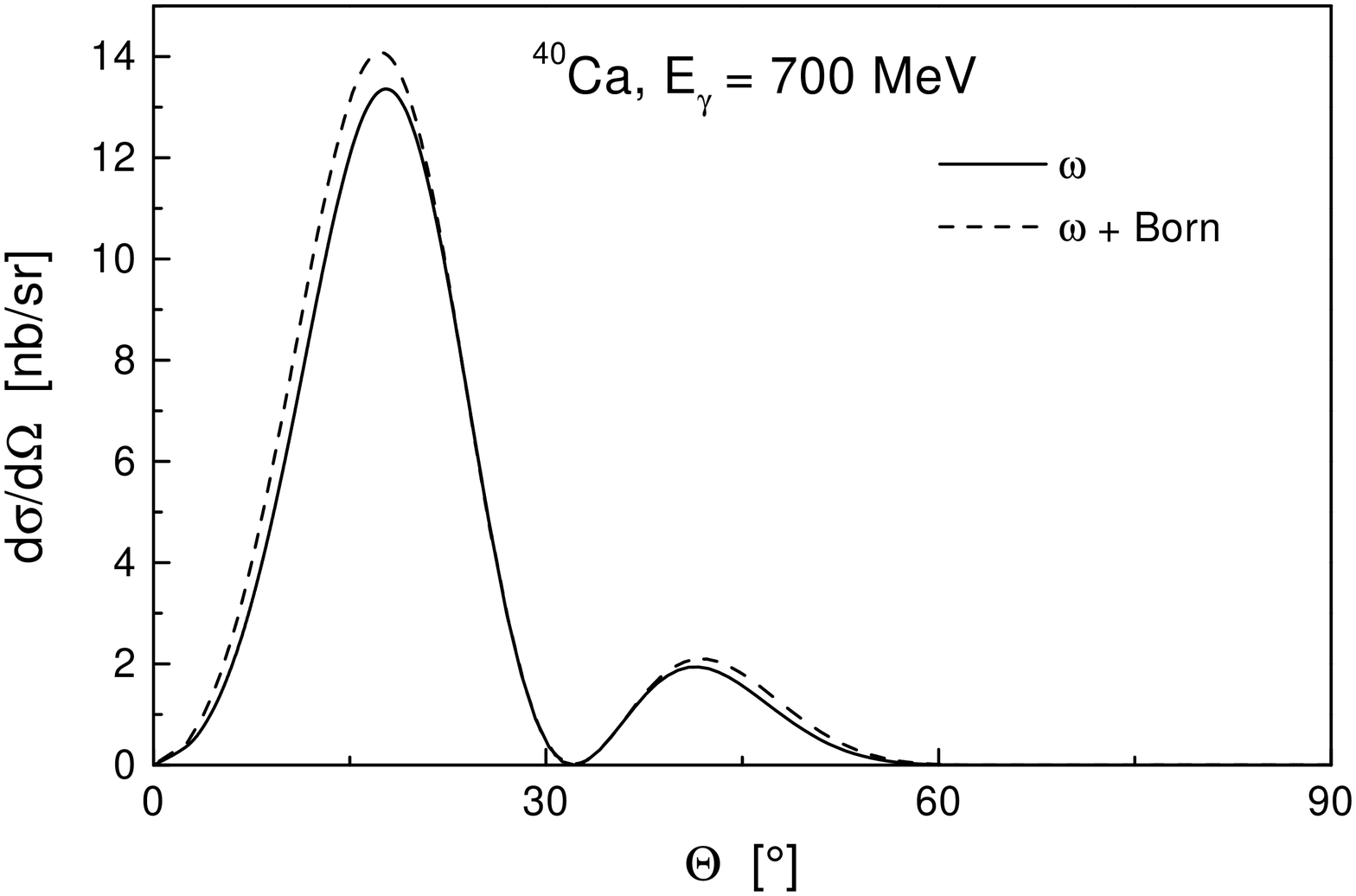} } 
  \caption{Differential cross section for the coherent photoproduction
    of $\eta$-mesons on $^{40}{\rm Ca}$ for a photon laboratory energy of 700
    MeV resulting from the omega graph and the nucleon Born terms. 
}
\label{dscaom} 
\end{figure}

In \fig{dsccom} we show the differential cross section for the
coherent photoproduction of $\eta$ mesons on $^{12}{\rm C}$ at a
photon laboratory energy of 650 MeV when only the omega graph is taken
into account. This can directly be compared to the results of
\cite{fix}. Our cross section is about 30 \% below the one given
there, which we attribute to the different off-shell behavior of the
production operator used in \cite{fix}.  In \fig{dscaom} we show the
corresponding cross section for $^{40}{\rm Ca}$ at a laboratory energy
of 700 MeV, which can be compared to the results in \cite{pieka}.  Our
cross section is about a factor of two lower than the one given there
for $^{40}{\rm Ca}$.  For $^{12}$C a cross section is found in
\cite{pieka} that is larger than ours by an order of magnitude. As
discussed in \cite{pieka}, this strong enhancement for $^{12}$C is due
to the special features of the tensor form factor, to which the
differential cross section is proportional in this study.  Since in
our approach the $\omega$-contribution is governed by the vector form
factor, we do not see such an enhancement.

\begin{figure}[!ht]
  \centerline{ \includegraphics[width=13cm]{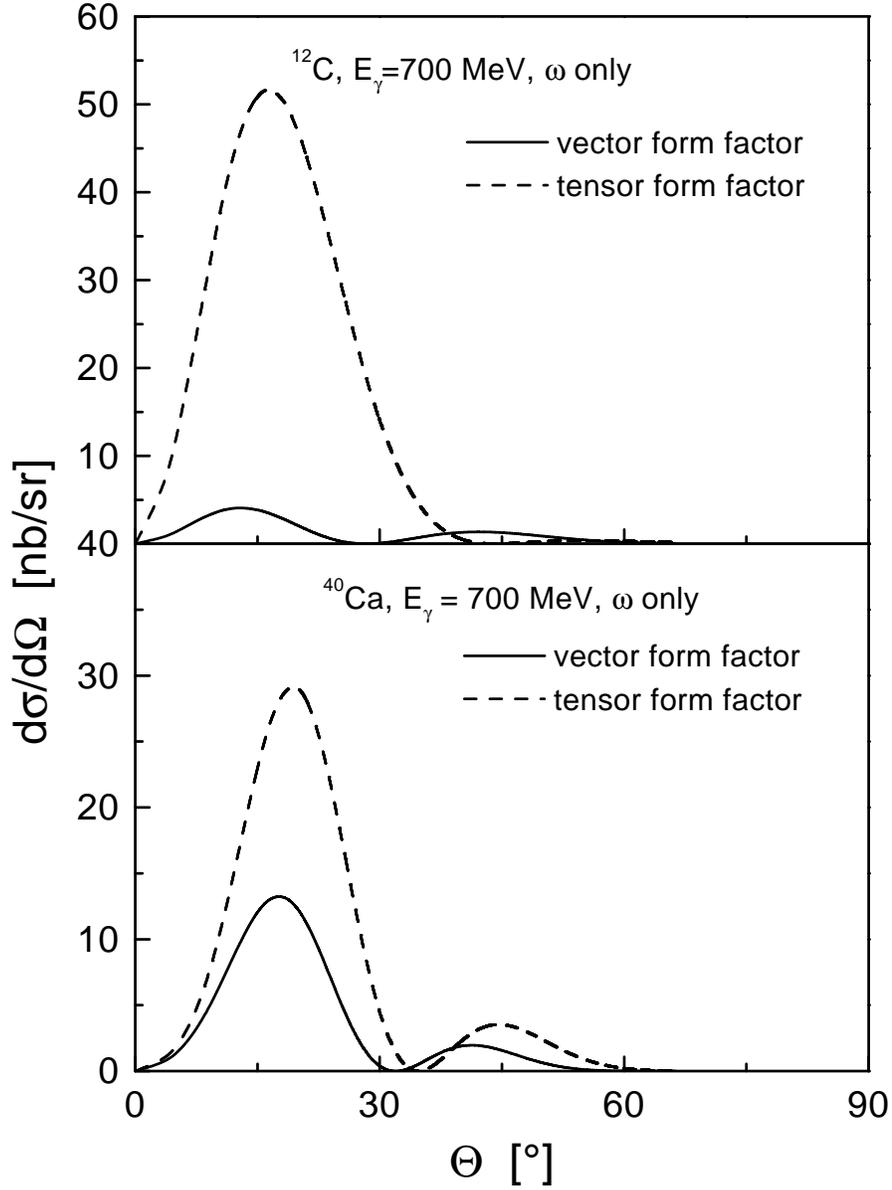} } 
  \caption{Differential cross section for the coherent photoproduction
    of $\eta$-mesons for a photon laboratory energy of 700 MeV
    resulting from the omega graph as in this work (vector form
    factor) and using the tensor form factor like in  \protect
    \cite{pieka} as described in the text.  }
\label{dstens} 
\end{figure}

In order to explicitly show that the large differences between our
results and those of \cite{pieka} are due to the different off-shell
behavior resulting from the replacement (\ref{ompieka}), we have
performed calculations for the omega graph using the approach from
\cite{pieka} together with our nuclear wave functions and coupling
constants.  The tensor term in \eq{ompieka} leads to an
$\omega$-contribution similar to \eq{omt2}, but with the vector form
factor replaced by \cite{mukop,pieka}:
\bea
F_v(Q) \to 2 m_N \frac{F_t(Q)}{Q}
\qc
\eea
where $F_t(Q)$ is the tensor form factor of the nuclear ground state:
\bea
F_t(Q) = 4 \pi\int dr\: r^2 j_1(Qr) \rho_t(r)
\qp
\eea
$\rho_t(r)$ is the tensor density defined in \eq{svt}. If the nucleon 
wave function is written as
\bea
\psi_{\alpha}(\vec{r}) = \left[
\begin{array}{r}
g_{a}(r) \cly_{J\ell}^M (\Omega_r) \\
if_{a}(r) \cly_{J\ell'}^M (\Omega_r)  
\end{array}
\right]
\quad {\rm with} \quad \alpha = (a,M), a=(n\ell J)
\qc
\eea
where $\cly_{J\ell}^M (\Omega)$ is the two component spin angle function,
$\rho_t(r)$ is given by:
\bea
\rho_t(r) = 2 \sum\limits_{a\:occ.}
\left(\frac{2J+1}{4\pi}\right) g_{a}(r)f_{a}(r) 
\qp
\eea
The results of this calculation are shown in \fig{dstens} for a photon
energy of 700 MeV in comparison to the results obtained by using our
form of the production operator in \eq{omgammu}.  
Since the two different operators (\ref{omgammu}) and (\ref{ompieka})
are equivalent on-shell, 
it is clear from \fig{dstens}
that the large differences between our results and \cite{pieka}
are indeed due to the different off-shell behavior of the production
operators. Note that the tensor
density, being linear in the small component $f_a$ of the nuclear wave
function, is very sensitive to the details of the nuclear structure
and to relativistic effects, in contrast to the vector density.

Also shown in \figs{dsccom} and \ref{dscaom} are the results of
calculations including in addition the nucleon Born terms. The
resulting curves differ very little from the results containing
only the omega graph. This is due to the small $\eta NN$ coupling
constant, and is in agreement with the findings of  \cite{fix} and
\cite{pieka}.

\subsection{The $N(1535)$}
\label{s11}

The \sss resonance strongly dominates the photoproduction of $\eta$
mesons on a single nucleon. All previous works on the coherent
photoproduction of $\eta$ mesons on nuclei found an almost complete
suppression of the \sss contribution to the coherent cross section, so
that the omega and the \ddd graphs give the largest contribution.  One
trivial reason for this suppression is the fact that the
$N(1535)p\gamma$ and the $N(1535)n\gamma$ coupling constant have about
the same size, but opposite sign (cf. Tab.~\ref{cc}), so that in
the coherent sum over all bound nucleons there is a large cancellation
between the proton and the neutron terms.  There is, however, another
source of suppression, which is due to the spin structure of the
$N(1535)$-amplitude \cite{bennhold}.  In order study this point,
we will now discuss the properties of the 
contribution of the
direct $N(1535)$-graph to the production operator.

In order to be able to separate local and non-local effects,
it is useful to first consider the leading non-relativistic terms in
the production amplitude. This is done by dropping the small
components of the nuclear wave functions and the intermediate
propagator. 
The term $\hat{T}_{S_{11}}$ defined in \eq{amplex} has the
explicit form:
\bea
\label{t110}
\hat{T}_{S_{11} } (E;\vec{x},\vec{y}) =
\frac{eg_{\gamma N S_{11}} g_{_{\eta NS_{11}}}}{2m_N}
G_{S_{11}} (E;\vec{x},\vec{y})
\gamma_5\:\sigma_{\mu \nu}\:k^\nu\varepsilon^\mu
\qp
\eea
In leading non-relativistic order this becomes a $2\times2$ matrix:
\bea
\label{t11}
\hat{T}_{S_{11} }^{(n.r.)} (E;\vec{x},\vec{y}) =
i\frac{eg_{\gamma N S_{11}} g_{_{\eta NS_{11}}}}{2m_N}
(E+m_{S_{11}})\:D(E;\vec{x},\vec{y})\:
E_\gamma \:\vec{\sigma} \vec{\varepsilon}\:
\qc
\eea
where 
\bea
D(E;\vec{x},\vec{y})=
\int \: \frac{d^3p}{(2\pi)^3} \:
            \frac{\rmi \rme^{\rmi \vec{p} ( \vec{x}-\vec{y})}}
            {E^2 - \vec{p}^2 - m_{S_{11}}^2+ i m_{S_{11}} \Gamma(s)}
\qp
\eea
The occurrence of the $\vec{\sigma} \vec{\varepsilon}$ term in
\eq{t11} is a direct consequence of the quantum numbers of the
$N(1535)$: Being an $S_{11}$ state, the \sss contributes predominantly
to the $E_{0^+}$ multipole, which is multiplied by $\vec{\sigma}
\vec{\varepsilon}$ in the CGLN-form for the elementary production
operator \cite{bennhold}. Thus the \sss leads in leading
non-relativistic order to a production operator that flips the spin of
the nucleon.

Now the question arises whether such a spin-flip term can contribute
to the coherent photoproduction. In the transition amplitude
(\ref{amplex}) matrix elements of the production operator are
evaluated between nuclear single-particle wave functions with the same
quantum numbers.  Since the total angular momentum is a good quantum
number, a spin-flip operator like the one in (\ref{t11}) can only
yield a contribution if the flip of the spin $m_s$ is compensated for
by a corresponding change in the orbital angular momentum component $m$.
In closed-shell nuclei such as $^{40}$Ca, this compensation is not
possible, since all $(m,m_s)$ states are occupied. Thus for such
nuclei, the \sss does not contribute to the coherent photoproduction
of $\eta$-mesons, at least in this non-relativistic picture, that
ignores spin-orbit effects.

For open-shell nuclei like $^{12}$C, however, the above argument no
longer holds. As will be shown below, in this case the contribution of
the \sss depends crucially on whether a local or a non-local treatment
is used. Only in the non-local case a spin-flip can be compensated by
a change of the orbital angular momentum. The \sss is therefore an
explicit probe for non-local effects, and its contribution will
necessarily be enhanced in a non-local approach, even in a
relativistic calculation.

In order to put these arguments on a formal basis, we write the
general form of the non-relativistic production operator, which is a
$2\times2$ matrix, in the following way:
\bea
\label{tlk}
\hat{T}^{(n.r.)}(E;\vec{x},\vec{y}) = L + i\vec{K}\vec{\sigma}
\qc
\eea
where $L=L(E;\vec{x},\vec{y})$ and
$\vec{K}=\vec{K}(E;\vec{x},\vec{y})$ stand for the spin-non-flip and
the spin-flip part of the production operator, respectively. The
non-local nuclear density matrix is in leading non-relativistic order
also a $2\times 2$ matrix:
\bea
\label{rho11}
\hat{\rho}_A^{(n.r.)}(\vec{x},\vec{y}) =
\sum\limits_{a\:occ.} 
g_{a}(x)g_{a}(y) \sum\limits_{M}
\cly_{J\ell}^M (\Omega_x) \otimes {\cly_{J\ell}^M}^+ (\Omega_y)
\qc
\eea
where $a=(n\ell J)$, $g_{a}$ is the radial wave function of the
bound nucleon and $\cly_{J\ell}^M (\Omega)$ is the two component spin
angle function.  For nuclei where all sub-shells (but not necessarily
all major shells) are completely occupied it can be shown that
\cite{negele}:
\bea
\label{rhononloc}
\hat{\rho}_A^{(n.r.)}(\vec{x},\vec{y})
&=& \alpha(\vec{x},\vec{y}) + \vec{\beta}(\vec{x},\vec{y}) \vec{\sigma}
\qc
\eea
with
\bea
\label{albet}
\alpha(\vec{x},\vec{y})&=& 
\sum\limits_{a\:occ.} 
\frac{g_{a}(x)g_{a}(y)}{4\pi} 
(J+\eh)P_\ell(\cos\theta) 
\nn\\
 \vec{\beta}(\vec{x},\vec{y}) &=&  i
\sum\limits_{a\:occ.} 
(-1)^{(\ell-J-\frac {1}{2})} \frac{g_{a}(x)g_{a}(y)}{4\pi} 
    P_\ell'(\cos\theta)
\left[{\hat{\vec{x}}} \times {\hat{\vec{y}}}\right]\: 
\qp
\eea
$\theta$ is the angle between $\vec{x}= x \hat{\vec{x}} $ and
$\vec{y}=y\hat{\vec{y}}$, and $P_\ell$ and $P_\ell'$ are the Legendre
polynomial of order $\ell$ and its derivative, respectively.  
Thus \eq{amplextr} 
for the transition amplitude
leads in the non-relativistic case to the formula:
\bea
\label{trnonrel}
T^{(\lambda)} &=&  \int d^3 x \: d^3 y \:
Tr\left[\hat{T}^{(n.r.)} (E;\vec{x},\vec{y}) 
\:\:\hat{\rho}_A^{(n.r.)}(\vec{x},\vec{y})\right] 
    \: \rme^{\rmi \vec{k} \vec{y}} \rme^{-\rmi \vec{q} \vec{x}}
\nn\\
&=& 
2\int d^3 x \: d^3 y \:
(L \alpha + i\vec{K} \vec{\beta} )
    \: \rme^{\rmi \vec{k} \vec{y}} \rme^{-\rmi \vec{q} \vec{x}}
\qc
\eea
where $L$, $\vec{K}$ from \eq{tlk} and $\alpha$, $\vec{\beta}$ from
\eq{rhononloc} are functions of $\vec{x}$ and $\vec{y}$. From its
definition in \eq{albet} it is clear, that
$\vec{\beta}(\vec{x},\vec{x})=0$. Thus the spin-flip part 
$\vec{K}\vec{\sigma}$ of
the production operator does not contribute in a local calculation
(cf.~\eq{locpos}).  However, in a non-local calculation the spin-flip
part does lead to a non-vanishing contribution via
$\vec{\beta}(\vec{x},\vec{y})\neq 0$ for $\vec{x}\neq\vec{y}$. The
non-spin-flip part $L$ contributes both locally and non-locally.

From the definition of $\vec{\beta}$ in \eq{albet} we can now read off
that the size of the non-local effects, arising from the $\vec{K}
\vec{\beta}$-term in \eq{trnonrel}, indeed depends on the details of
nuclear structure: For a completely occupied shell with given $\ell$
the two orbitals with different $J$ contribute to $\vec{\beta}$ with a
different sign.  If spin-orbit effects are neglected, i.e.~if the
radial wave functions $g_a$ are the same for both orbitals,
their contributions to $\vec{\beta}$ cancel exactly, in agreement with
the argument we gave above.  For a system like $^{12}{\rm C}$,
however, where only the $1p\drh$ orbital is occupied and the $1p\eh$
orbital is empty, the contribution of the $1p\drh$ orbital is not
cancelled.  Since we are using a relativistic equation of motion for
the bound states (\eq{nuceom}), our wave functions contain spin-orbit
effects, but the radial wave functions of two orbitals within one
shell are still rather similar. The cancellation in $\vec{\beta}$ in
the case of a closed-shell nucleus is thus not complete, but this term
is still strongly reduced for $^{40}{\rm Ca}$ as compared to
$^{12}{\rm C}$.  Thus there will be a non-local contribution from the
\sss resonance, which is enhanced for $^{12}{\rm C}$ as compared to
$^{40}{\rm Ca}$.

\begin{figure}[!ht]
  \centerline{ \includegraphics[width=13cm]{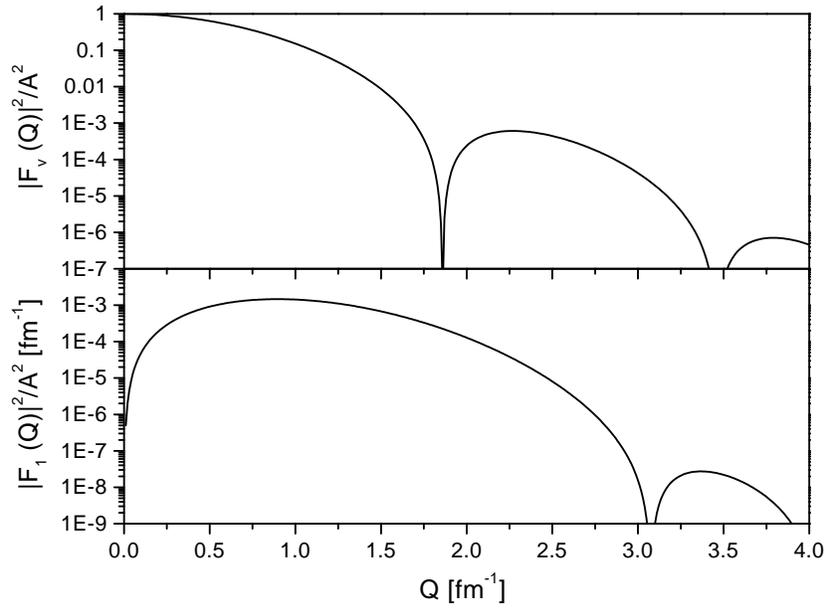} } 
  \caption{The vector form factor $F_v$ and the modified form factor $F_1$ 
    for $^{12}{\rm C}$ as defined in the text.  }
\label{ffcc} 
\end{figure}

Besides this dependence on the nuclear structure non-local effects
also lead to a different angular dependence than purely local
contributions.  The reason is that the non-local parts admix higher
multipole components to the transition matrix element.  This can be
seen by calculating non-local corrections to the local approximation
in \eq{locpos}.  This is done by making a Taylor expansion of the
production operator $\hat{T} ( E;\vec{p},\vec{k},\vec{q})$ around a
fixed momentum $\vec{p}=\vec{p}_o$ and considering the first order
correction to the local approximation.  One  finds for the 
transition amplitude (cf.~\eq{amplexloc})
\bea
\label{locposcorr}
T^{(\lambda)} 
 &=& Tr \left[
         \hat{T}^{(\lambda)} ( E;\vec{p}_o,\vec{k},\vec{q}) \:\:
         \hat{\rho}_A(\vec{k}-\vec{q}) \right] +
Tr \left[\left(
         \grad_{p}
         \hat{T}^{(\lambda)} ( E;\vec{p},\vec{k},\vec{q})
         \right)_{\vec{p}=\vec{p}_o} \cdot
         \hat{\vec{\rho}}_A \right]
+ \ldots
\qp
\eea
Taking $\vec{p}_o = -\eh(\vec{k} - \vec{q})$, { which is similar to
  the standard choice discussed in \cite{pions}}, $\hat{\vec{\rho}}_A$
is given by the difference of two dyadic products:

\bea
\hat{\vec{\rho}}_A=
\frac{\rm i}{2}\sum\limits_{\alpha \:occ.} \: \int d^3 x \left(
  \psi_\alpha \otimes (\grad \overline{\psi}_\alpha)  -   
  (\grad \psi_\alpha)\otimes\overline{\psi}_\alpha 
\right) {\rm e}^{i(\vec{k}-\vec{q})\vec{x}}
\qp
\eea
The spatial derivatives in the first order
correction change the angular momentum structure of the matrix
element.  
Inserting the non-relativistic production operator (\ref{tlk}) into 
\eq{locposcorr}, the spin-flip part
$\vec{K}\vec{\sigma}$ leads to a first order non-local
correction that involves a modified nuclear form factor,
which for the case of $^{12}$C has the  form:
\bea
 F_1(Q)=\:4 \pi
         \int r\:dr\: j_1(Qr)  g_{p\drh }^2(r) 
\qc
\eea        
where $g_{p\drh }$ is the ${1p\drh }$ radial wave function
and $Q = \mid\vec{k}-\vec{q}\mid$ is the transfered momentum.
 Due to the
spatial derivatives in \eq{locposcorr} a spherical Bessel function of
first order appears, in contrast to the vector form factor
in \eq{rhofou}, which involves
spherical Bessel function of order 0.
Thus it is apparent, that non-local effects introduce higher
multipolarities, in this case a dipole component.  In \fig{ffcc} $F_1$
is shown in comparison to the vector form factor  for
$^{12}{\rm C}$.  $F_1$ shows a very different dependence on the
transfered momentum than the vector form factor.  
We can conclude from this difference
that the
contribution of the $N(1535)$, which is in leading non-relativistic
order purely non-local, has a different angular dependence than the
omega term, which is proportional to the vector form factor $F_v$.

\begin{figure}[!ht]
  \centerline{ \includegraphics[width=13cm]{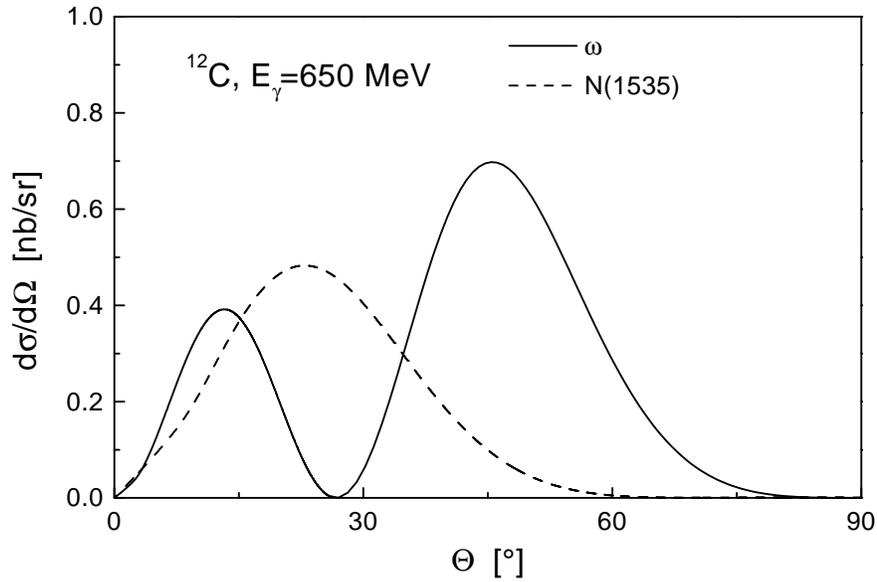} } 
  \caption{
        $N(1535)$-contribution to the coherent photoproduction of
    $\eta$-mesons on $^{12}{\rm C}$ at a photon laboratory energy of
    650 MeV in comparison to the omega contribution.  
}
\label{dsccom11} 
\end{figure}

In \fig{dsccom11} we show the contribution of the direct
$N(1535)$-graph to the coherent cross section on $^{12}{\rm C}$ in a
fully relativistic, non-local calculation in comparison to the
$\omega$-contribution.  The $N(1535)$-contribution is comparable to
the omega term, but shows a very different angular dependence, which
is in qualitative agreement with the different $Q$-dependence of the
higher order form factor from \fig{ffcc} (A photon energy $E_\gamma$=
650 MeV and a scattering angle between 0 and 90$^o$ correspond to a
momentum transfer between about 1.5 fm$^{-1}$ and 3.5 fm$^{-1}$).

\begin{figure}[!ht]
  \centerline{ \includegraphics[width=13cm]{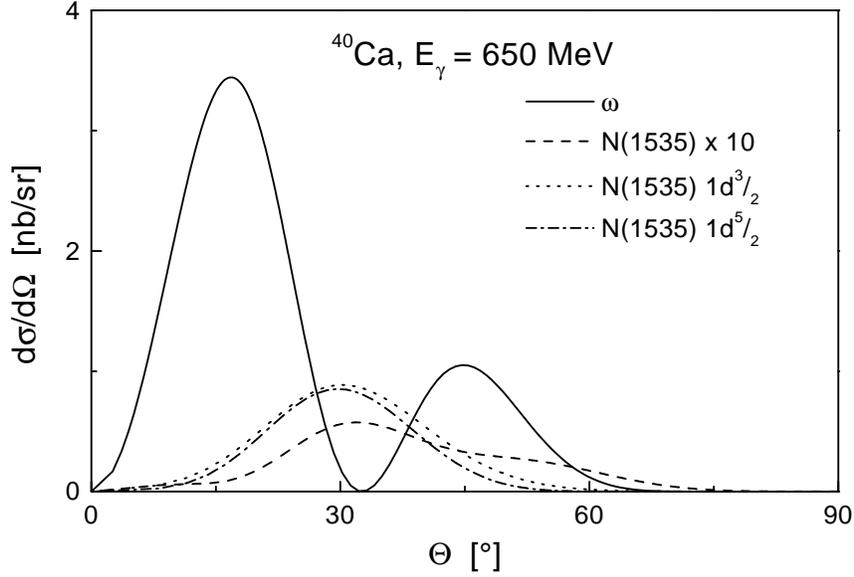} } 
  \caption{$N(1535)$-contribution to the coherent photoproduction of
    $\eta$-mesons on $^{40}{\rm Ca}$ at a photon laboratory energy of
    650 MeV. The $N(1535)$-contribution is 
multiplied by 10. Also shown are the results for the
    $N(1535)$-contribution if only the $1d\drh$ and the $1d\fh$
    orbital are taken into account.  }
\label{dscaom11} 
\end{figure}

In \fig{dscaom11} we compare the \sss and the omega contributions for
$^{40}{\rm Ca}$. For this nucleus the \sss yields only a small
contribution to the coherent cross section.  We thus find that the
size of non-local corrections relative to the leading local term is
smaller for the closed-shell nucleus $^{40}{\rm Ca}$.  In order to
show that this is indeed due to a cancellation between the orbitals
within one shell, we also show in \fig{dscaom11} the contribution of
the $1d\drh$ and the $1d\fh$-orbital to the \sss term separately.  The
total \sss contribution is an order of magnitude below the one of the
individual orbitals, in agreement with the argument we gave above.
Thus,  the relative importance of the \sss in the coherent
photoproduction is directly affected by the shell structure of the
target nucleus. The relative strength shows strong variations when
going from closed-shell to closed sub-shell nuclei.

{ 
  
  The importance of non-local effects in the coherent photoproduction
  of $\eta$-mesons is a result of the suppression of the \sss in a
  local calculation for this process.  Such a suppression does not
  occur in the inclusive $\eta$-photoproduction (see
  e.g.\cite{carrasco}). In the coherent photoproduction of pions
  non-local effects are also of minor importance since the $\Delta$
  yields the dominant contribution in a local as well as in a
  non-local calculation. Thus only the coherent photoproduction of
  $\eta$-mesons shows a dependence on non-local effects that is strong
  enough to study these effects in detail.}

\subsection{The $N(1520)$}
\label{d13}

Being a spin-$\frac{3}{2}$ particle, the \ddd leads to more
complicated expressions for the production operator. We take the
spin-$\frac{3}{2}$ propagator to be:
\bea
G_{D_{13}}^{\mu\nu}(p) &=& \rmi \:
 \frac{\bsl p + m_\Delta }
{p^2-m_{D_{13}}^{2} + \rmi m_{D_{13}}\:\Gamma} \:
\Lambda^{\mu\nu}
\label{d13prop}
\qc
\eea
with $\Gamma$=120 MeV \cite{fix} and
\bea
\Lambda^{\mu\nu}\:=\:
\left( g^{\mu \nu} -
\frac{1}{3}\gamma^{\mu}\gamma^{\nu} -
\frac{2}{3m_{D_{13}}^{2}}p^{\mu}p^{\nu} - 
\frac{1}{3m_{D_{13}}} 
(\gamma^{\mu} p^{\nu} - p^{\mu} \gamma^{\nu}) \right) 
\label{d13prop32}
\qp
\eea

The three-momenta $p_i$ in the last term in \eq{d13prop32} are treated
exactly by replacing them with gradient operators in position space
and inserting derivatives of wave functions wherever it is necessary.
The spatial components of the $p^{\mu}p^{\nu}$ term in \eq{d13prop32}
lead to second derivatives, the exact evaluation of
which leads to numerical complications. We therefore make the
replacement: 
\bea
\label{locpp}
p_i\:p_j  \:\to\: k_i\:k_j
\qc
\eea
where $\vec{k}$ is the photon momentum. This is equivalent to setting
the momentum of the incoming nucleon for this particular term to zero,
corresponding to a local approximation.  

This is unfortunately not the only complication arising for the \ddd
resonance. From \eq{lint} one sees that the second coupling of this
resonance to the photon contains the momentum of the nucleon.  Thus
for the second coupling many more terms arise that contain higher
powers of derivatives of wave functions, which would require a rather
involved numerical treatment.

\begin{figure}[!ht] 
  \centerline{ \includegraphics[width=13cm]{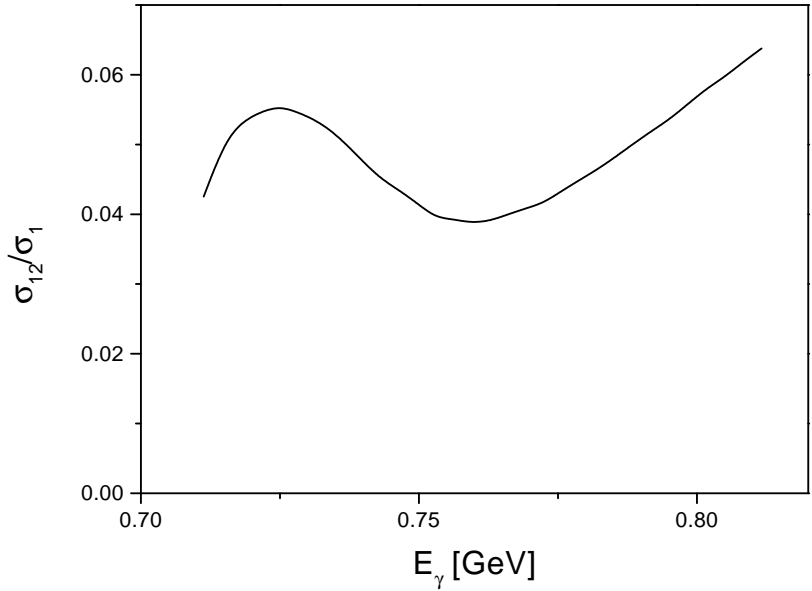} } 
  \caption{
    The ratio of the \ddd contributions to the isoscalar cross
    section of the elementary photoproduction of $\eta$ mesons as a
    function of the photon energy in the laboratory frame.
    $\sigma_{12}$ denotes the entire \ddd contribution, including both
    types of couplings to the photon, while $\sigma_{1}$ is the \ddd
    contribution when only the first kind of coupling is used.
    }\label{ratio}
\end{figure}

{
  
  For this reason our calculations are restricted to the first kind of
  coupling of the \ddd to the photon. In order to interpret our
  results correctly, one first has to consider the role of the two
  different couplings to the photon in the elementary photoproduction.
  For this purpose we show in \fig{ratio} the ratio of the isoscalar
  $N(1520)$-contribution to the elementary photoproduction of $\eta$
  mesons when both couplings are taken into account and when only the
  first one is used.  The value of this ratio of about 1/20 shows,
  that using only the first kind of coupling yields a contribution
  which is about a factor of 20 larger than the one resulting from the
  inclusion of both kinds of couplings.  Thus the total
  $N(1520)$-contribution to the photoproduction of $\eta$-mesons is
  the result of a strong cancellation between the two coupling types.
  Since we only include the first kind of photon coupling of the
  $N(1520)$, we have to account for this cancellation.  This is done
  in an approximate way by using coupling constants that are rescaled
  by a factor of $\sqrt{1/20}$.  This procedure can only yield a rough
  estimate, but it allows us to draw qualitative conclusions about the
  role of the \ddd in our approach.

}

\begin{figure}[!ht]
  \centerline{ \includegraphics[width=13cm]{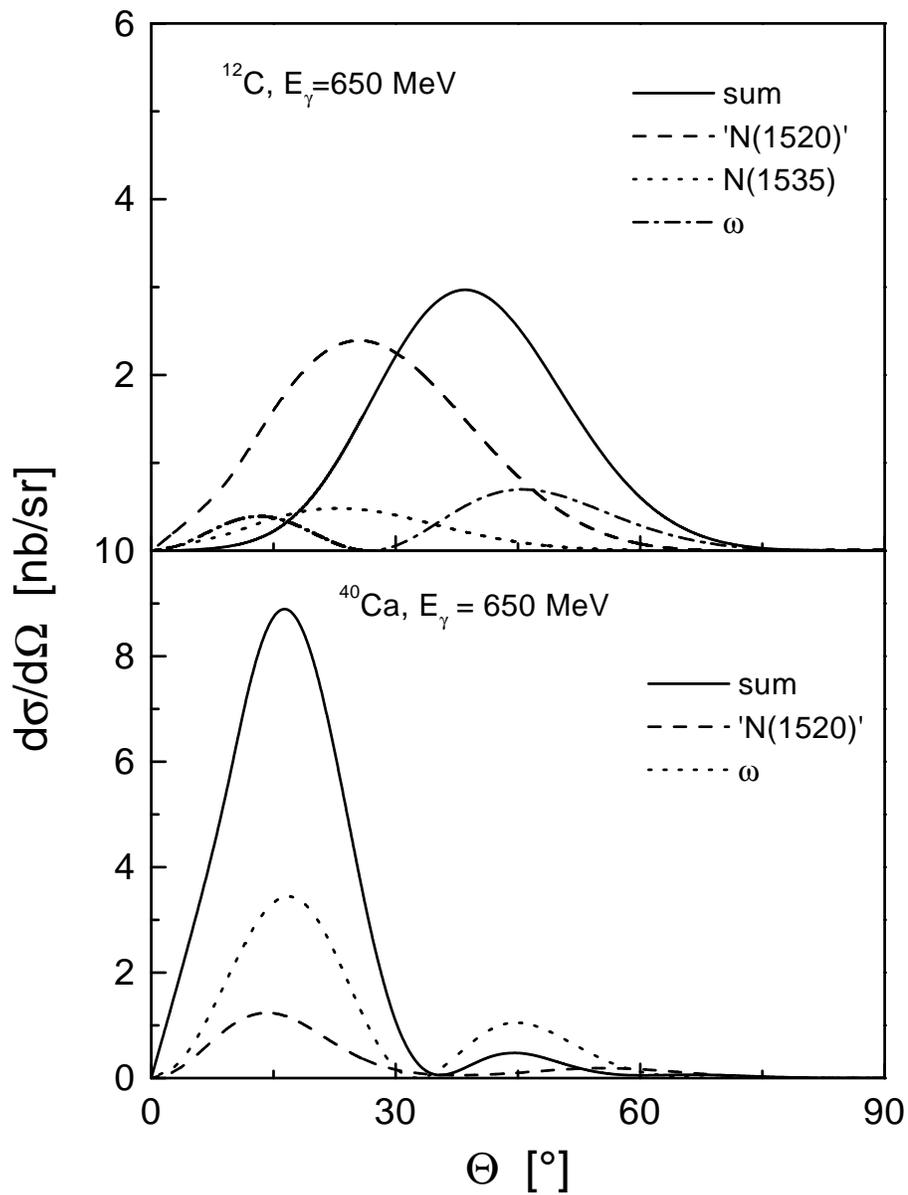} } 
  \caption{Complete
    differential cross section for the coherent production of
    $\eta$-mesons on $^{12}{\rm C}$ and $^{40}{\rm Ca}$ together with
    the separate contributions and the estimate of the \ddd term
    at a photon laboratory energy of 650 MeV.  }
\label{dscont}  
\end{figure}

In the local calculations of \cite{fix,pieka}, the \ddd contribution
is smaller than the one from the omega.  In the upper part of
\fig{dscont} the contribution of the direct \ddd graph for $^{12}{\rm
  C}$ in our model is shown, together with the other contributions
that we discussed above. The \ddd yields, in the approximation
described above, the largest contribution. From the angular dependence
we can conclude that this is mainly due to non-local effects, just as
in the case of the \sss resonance.  This is in agreement with the fact
that the \ddd appears in the elementary photoproduction of
$\eta$-mesons in the $E_{2^-}$ and the $M_{2^-}$ multipole, which
mainly contribute to the spin-flip part of the CGLN-form of the
amplitude \cite{bennhold}.  That the \ddd contributes even more than
the \sss is due to the fact that the isospin averaging leads to a
strong cancellation of proton and neutron contributions in the case of
the \sss because of the numerical values of its couplings to the
photon.  This can be seen by extracting the ratio of isoscalar and
isovector $N(1535)N\gamma$ and $N(1520)N\gamma$ coupling constants
from the values given in Table \ref{cc}.  This ratio is given by
$g({T=0})/g({T=1})=(g_p+g_n)/(g_p-g_n)$. For the \sss we find
$g({T=0})/g({T=1})=$0.08, which means that the $N(1535)N\gamma$
coupling is strongly dominated by an isovector coupling.  For the \ddd
we find for its two kinds of coupling to the photon
$g({T=0})/g({T=1})=$0.7 and 1.3, respectively, so that isoscalar and
isovector couplings are of comparable size for this resonance. For
$N=Z$ nuclei like $^{12}{\rm C}$ and $^{40}{\rm Ca}$ with total
isospin $T=0$, the coherent process is almost completely determined by
the isoscalar coupling.  This isospin selection strongly suppresses
the \sss contribution, while it affects the \ddd resonance much less.

In the lower part of \fig{dscont} we show the corresponding results
for $^{40}{\rm Ca}$. From the non-local character of the \ddd
contribution in the case of $^{12}{\rm C}$ it is clear that it
contributes less for $^{40}{\rm Ca}$ because of the suppression of
non-local effects for this nucleus; it turns out to be smaller than
the $\omega$-contribution.  Thus the differential cross section is for
$^{40}{\rm Ca}$ still dominated by local contributions, in contrast to
the case of $^{12}{\rm C}$, where the size and the shape of the
differential cross section is governed by the large non-local
contributions of the \sss and the the $N(1520)$.  It must of course be
kept in mind, that the \ddd contribution we find is only an estimate.

\subsection{The complete cross section and the eta-nucleus interaction}
\label{compl}

After having studied the properties of the single contributions to the
coherent photoproduction of $\eta$ mesons we now discuss the complete
cross section. We 
have
performed calculations including the omega graph,
the nucleon Born terms, direct and exchange graphs for the \sss and
the direct \ddd graph, treated approximately as described in
\msc{d13}. In \fig{dscont} the resulting differential cross section
for $^{12}{\rm C}$ and $^{40}{\rm Ca}$ at $E_\gamma$=0.65 GeV is
shown, together with the single contributions.  For $^{12}{\rm C}$ we
find a cross section which is at this energy about a factor of three
larger than the one in \cite{fix} and shows, as has been discussed
above, a very different angular dependence. Note that for both nuclei
the interference between the local $\omega$-term and the non-local
resonance contributions can be both, destructive or constructive,
depending on the angle. The reason is that the vector form factor
(\eq{rhofou}), and therefore the $\omega$-amplitude, changes its sign
at certain momentum transfers corresponding to the minima in the 
differential cross
section in Figs.~\ref{dsccom} and \ref{dscaom}.  The non-local
contributions from the \sss and \ddd, that are added coherently to the
omega amplitude, do not show such a change in sign, which leads to
the interference pattern seen in \fig{dscont}.

\begin{figure}[!ht]
  \centerline{ \includegraphics[width=13cm]{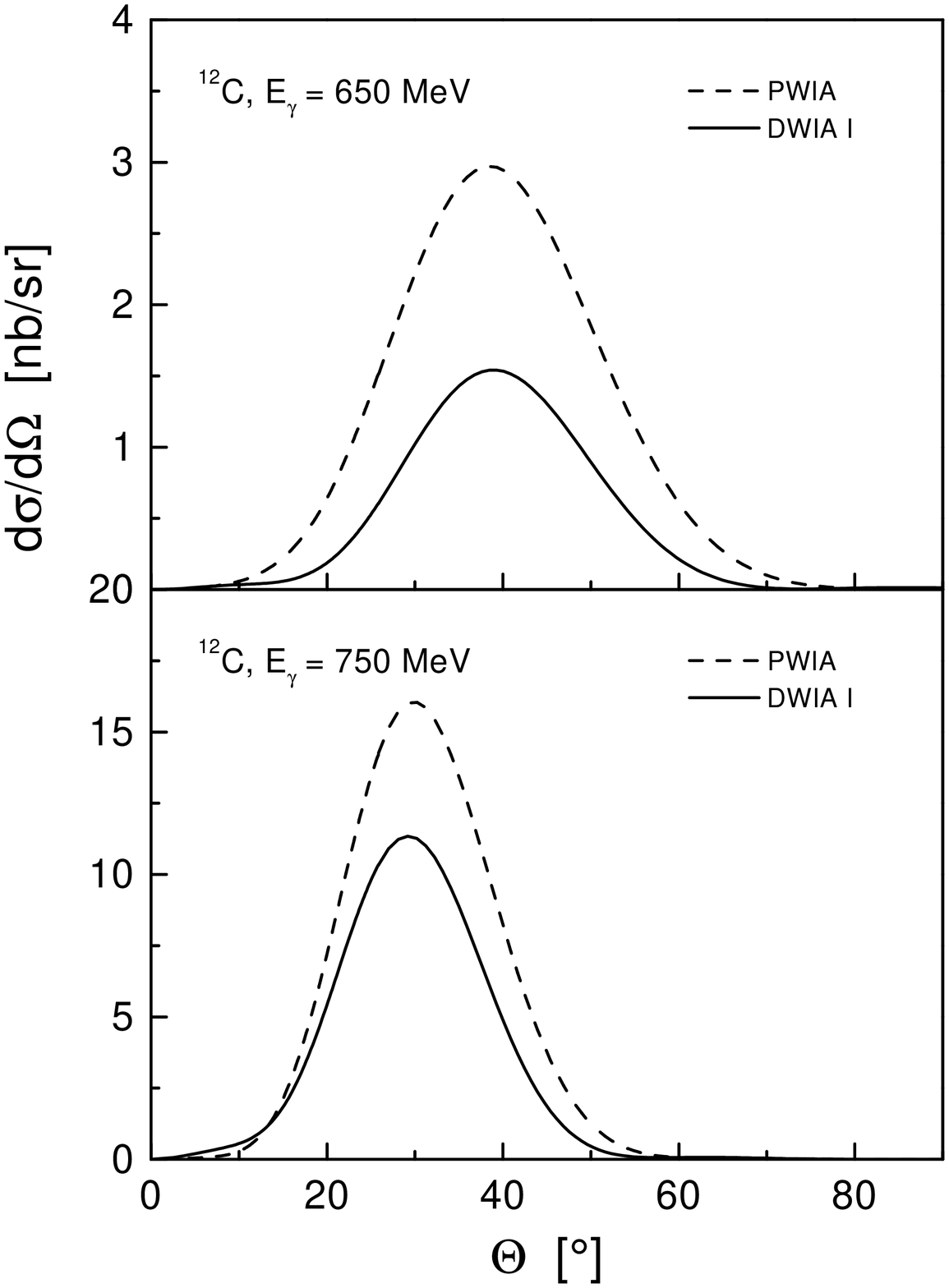} } 
  \caption{
    Differential cross section for the coherent production of
    $\eta$-mesons on $^{12}{\rm C}$ for two 
different photon laboratory energies in PWIA
    and in DWIA I.  }
\label{dsccdw}  
\end{figure}

\begin{figure}[!ht]
  \centerline{ \includegraphics[width=13cm]{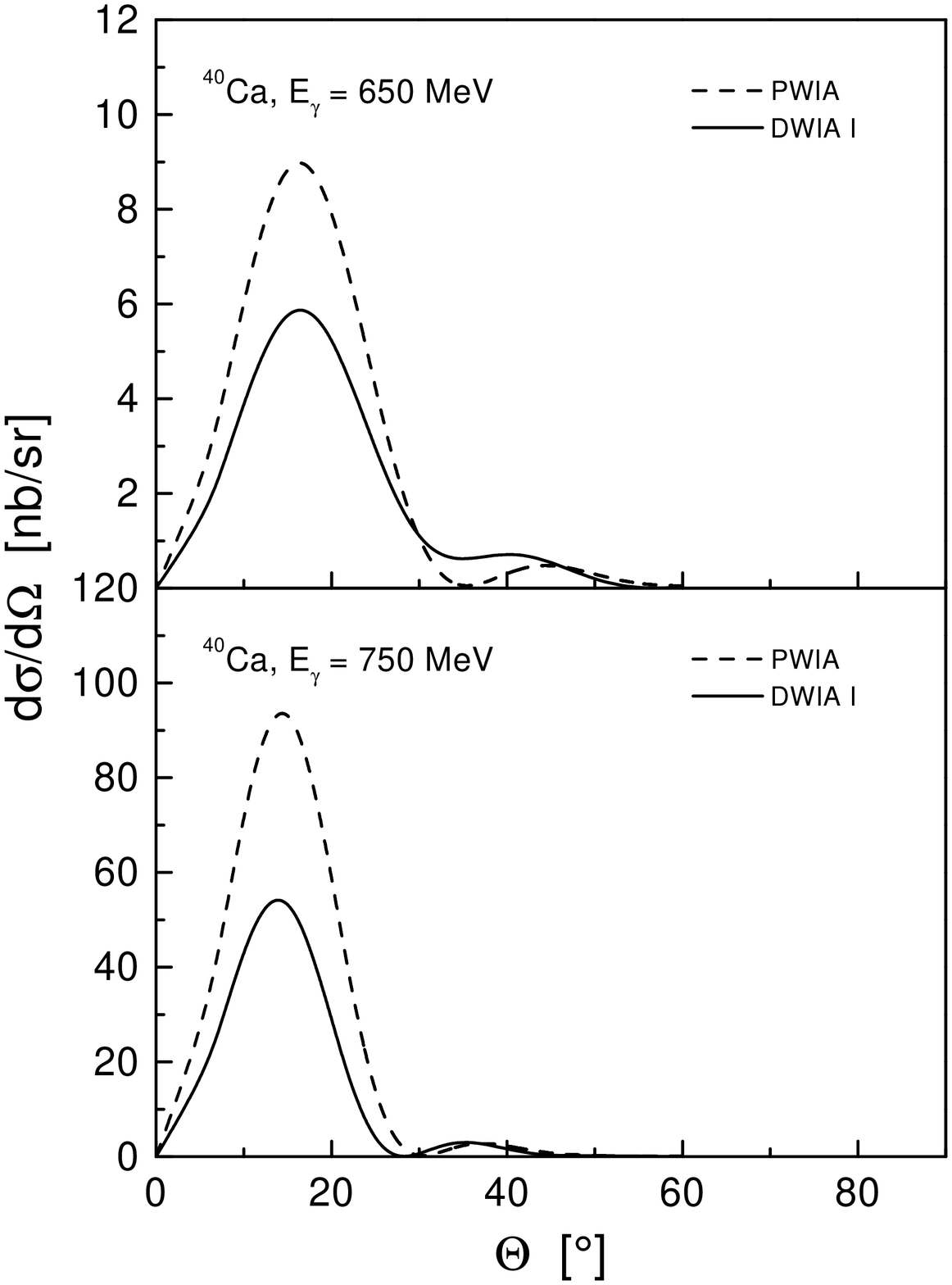} } 
  \caption{
    Differential cross section for the coherent production of
    $\eta$-mesons on $^{40}{\rm Ca}$ for two different photon
    laboratory energies in PWIA and in DWIA I.  }
\label{dscadw}  
\end{figure}

In \fig{dsccdw} we show the differential cross section for $^{12}{\rm
  C}$ for a photon energy of 650 and 750 MeV for a PWIA as well as a
DWIA calculation employing the optical potential I which is taken from
\cite{pieka}.  At these energies, an $\eta$ nucleus optical potential
mainly leads to an overall decrease of the cross section.  In
\fig{dscadw} the corresponding results for $^{40}{\rm Ca}$ are shown.

\begin{figure}[!ht]
  \centerline{ \includegraphics[width=13cm]{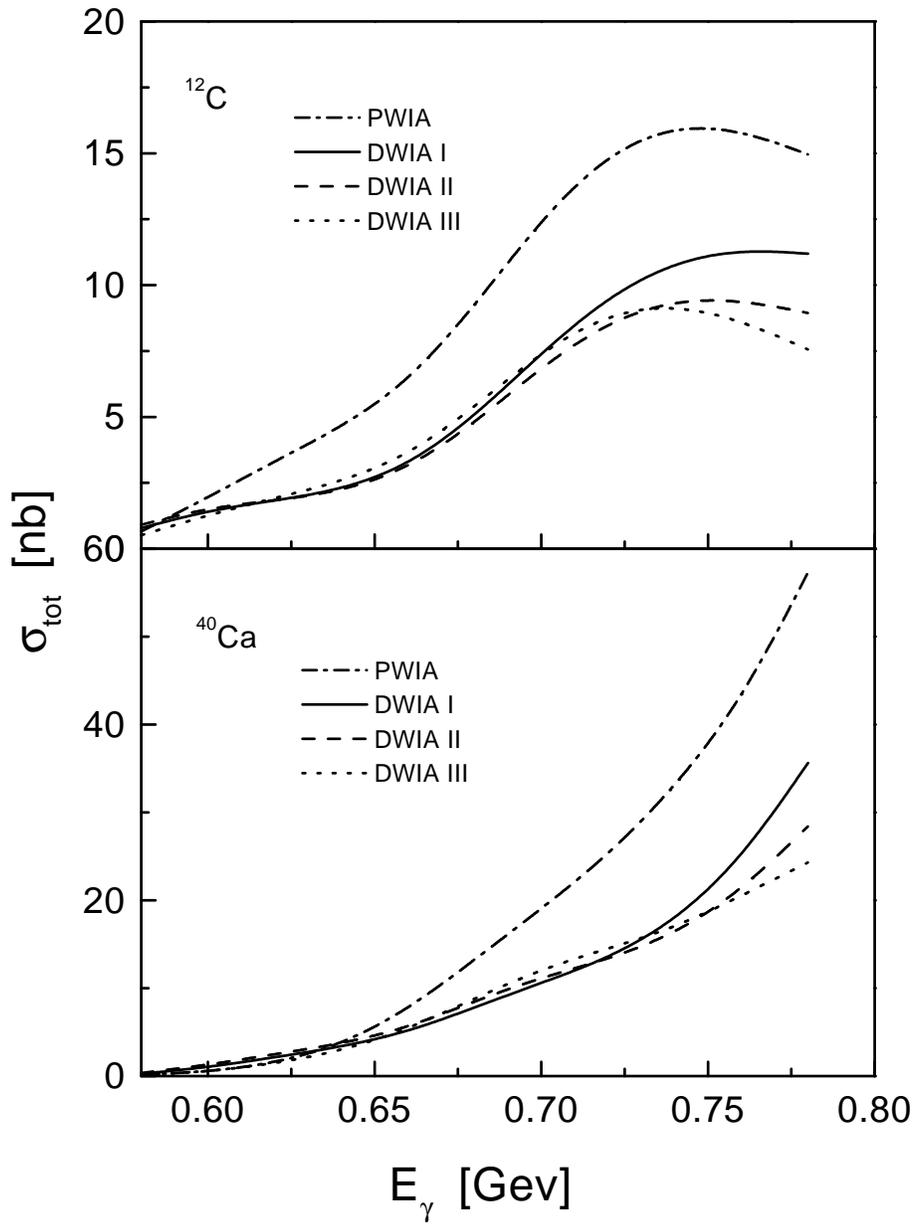} } 
  \caption{
    Total cross section for the coherent production of $\eta$-mesons
    on $^{12}{\rm C}$ and $^{40}{\rm Ca}$ in PWIA and in DWIA using
    three different optical potentials as a function
    of the photon laboratory energy.  }
\label{ssdw}  
\end{figure}

{ In \fig{ssdw} we finally show total cross sections for $^{12}{\rm
    C}$ and $^{40}{\rm Ca}$ as a function of the photon energy in PWIA
  and DWIA using the optical potentials I \cite{pieka}, II
  \cite{carrasco} and III \cite{effe}.  Due to the different
  importance of the non-local resonance contributions, the energy
  dependence of the cross section is not the same for the two nuclei,
  and the effect of including an optical potential is slightly
  different.  If only the local $\omega$-contribution is included,
  both the energy dependence of the total cross section and the effect
  of an optical potential are very similar for $^{12}{\rm C}$ and
  $^{40}{\rm Ca}$.

One sees from \fig{ssdw} that the inclusion of an optical potential
mainly leads to a decrease of the cross section.  All three optical
potentials lead to similar results at lower energies, while at higher
energies increasing differences are visible.
Setting the real part of the optical potential III equal to zero
is not a strong assumption, since the real part of potential I and II
has only very little influence on the cross section beyond 650 MeV.
}

\begin{figure}[!ht]
  \centerline{ \includegraphics[width=13cm]{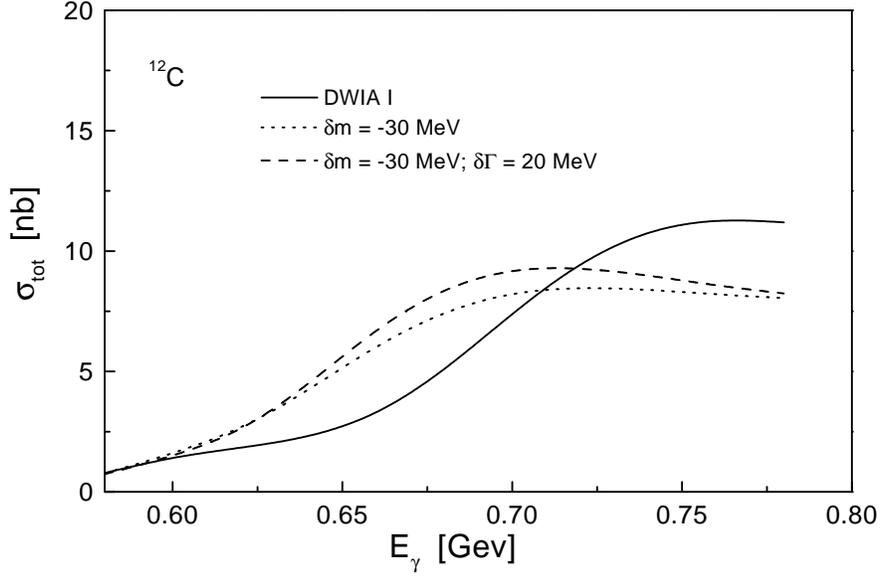} } 
  \caption{
    The effect of a medium modification of the $N(1535)$ as described
    in the text. Displayed is a calculation where the mass of the
    $N(1535)$ is changed, as well as one with an additionally modified
    width, in comparison to a calculation employing the free values.
}
\label{sgim}  
\end{figure}

{ 
  
  In \cite{carrasco} the properties of the \sss in nuclear matter are
  discussed. Using the results of this work we studied the
  influence of a modification of the $N(1535)$-properties in the
  nuclear medium on our results.  Employing the model developed in
  \cite{chiang}, a broadening of the $N(1535)$ by about 30 MeV at
  normal nuclear density is found in \cite{carrasco}. For the real
  part of the $N(1535)$ self-energy a potential $V=V_o \rho/\rho_o$
  with $V_o$=-50MeV is assumed.  Analogous to the procedure used in
  \cite{pions} for the $\Delta$-resonance, we include this effectively
  by changing the mass and the width of the $N(1535)$ by $\delta
  m$=-30 MeV and $\delta \Gamma$=20 MeV, respectively. The result of
  such a calculation is shown in \fig{sgim}. A change of the mass and
  the width of the $N(1535)$ leads to a shift of strength towards
  lower photon energies. Due to the smallness of the
  $N(1535)$-contribution for $^{40}{\rm Ca}$, the modifications of the
  $N(1535)$-properties do not lead to a significant effect for this
  nucleus.}

An experiment has been performed at MAMI, which is currently being
analyzed, where the photoproduction of $\eta$ mesons on $^{4}{\rm He}$
has been measured. The analysis of the experimental data might reveal
a coherent signal \cite{bernd2}, which could in principle lead to the
first data for this process. Although we are aware, that a mean field
approximation is only of limited validity for $^{4}{\rm He}$, we have
performed calculations for this nucleus.  We find that the form factor
of $^{4}{\rm He}$ is reasonably well reproduced for the parameters for
the potentials in \eq{nucpot} also given in Tab.~\ref{nucpotp}. The
binding energies, however, come out too large.  This is a known effect
in mean field calculations, related to explicit contributions from
many-body correlations \cite{schia}.

\begin{figure}[!ht]
  \centerline{ \includegraphics[width=13cm]{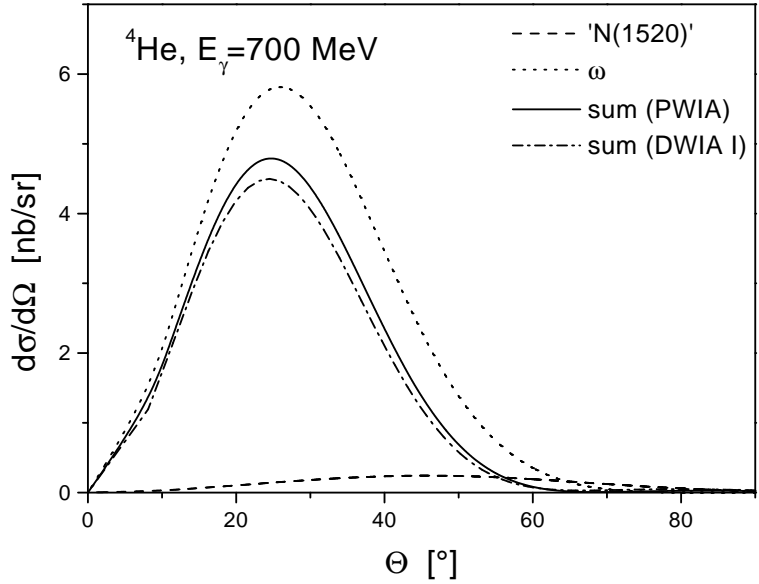} } 
  \caption{
    Differential cross section for the coherent production of $\eta$
    mesons on $^{4}{\rm He}$ at a photon laboratory energy of 700 MeV in
    a PWIA and a DWIA calculation.  }
\label{he1}  
\end{figure}

In Fig.~\ref{he1} we show the complete differential cross section for
$^{4}{\rm He}$ as well as the $\omega$-contribution and the
contribution from the \ddd resonance at a photon laboratory energy of
700 MeV.  The cross section in \fig{he1} is about a factor of two
below the one in \cite{fix} and a factor of three smaller than the one
given in \cite{pieka}.  The \sss contribution is negligible. 
In the approximation described above the \ddd
contribution is visible, but much smaller than the
$\omega$-contribution. We thus find that the coherent $\eta$
production on $^{4}{\rm He}$ is dominated by the local $\omega$-term.
This is easy to understand since $^{4}{\rm He}$ is a closed shell
nucleus, so that non-local effects are relatively smaller than in the
case of $^{12}{\rm C}$ (cf. \msc{s11}).  Due to the limited
applicability of the mean-field approximation to $^{4}{\rm He}$, this
result can only be considered as a first estimate. The suppression of
non-local effects, however, is independent from this uncertainty,
since it is a result of the quantum numbers of the $^{4}{\rm He}$ wave
functions.  Also shown in \fig{he1} is the result of a DWIA
calculation. The $\eta$-nucleus interaction only leads to a small
decrease of the cross section. This is due to the small absorption at
a light nucleus like $^{4}{\rm He}$.

\section{Summary and  Conclusions}
\label{sum} 

We have calculated differential and total cross sections for the
coherent photoproduction of $\eta$ mesons on $^{12}{\rm C}$ and $^{40}{\rm Ca}$ in
a relativistic, non-local model using the impulse approximation. This
model has previously been applied to the coherent photoproduction of
pions on nuclei, where good agreement with the experimental data was
found.  Previous calculations for the coherent production of $\eta$
mesons have used a local approximation and have found a strong
suppression of the \sss contribution, which dominates the elementary
process.

The two main differences between the recent studies
\cite{fix,pieka} and our model are the different off-shell
extrapolations used for the elementary production operator, and the
fact that our calculation contains non-local effects.

We have found a strong dependence of the $\omega$-contribution on the
off-shell extrapolation of the production operator, especially in
comparison to the approach of \cite{pieka}. We have also found that
non-local effects can lead to a sizable contribution of the $N(1535)$,
which shows a different angular dependence than the local
$\omega$-contribution. These non-local effects have been shown to
depend on the shell structure of the nucleus such that they are small
for a closed-shell system like $^{40}{\rm Ca}$, while being strongly
enhanced for the open-shell nucleus $^{12}{\rm C}$. The \sss is a
sensitive probe for these non-local effects, { which is a special
  feature of the coherent photoproduction of $\eta$-mesons.}

We have given an estimate for the contribution of the \ddd resonance
to the coherent production of $\eta$-mesons.  This estimate indicates
that the contribution of this resonance also contains non-local
effects. In the case of $^{12}{\rm C}$, it led to a \ddd contribution
which is even larger than the $\omega$-term. For $^{40}{\rm Ca}$,
where non-local effects are smaller, our estimate still yields a
sizeable $N(1520)$-contribution, which is, however, smaller than the
omega term.

The large resonance contributions to the coherent production on
$^{12}{\rm C}$ lead to a resonant behavior of the total cross section
for this nucleus. For $^{40}{\rm Ca}$, where the resonances contribute
much less due to the relative suppression of non-local effects, the
total cross section does not show a resonant behavior. This is in
contrast to the coherent production of pions, where the shape of the
total cross section for $^{12}{\rm C}$ and $^{40}{\rm Ca}$ is very
similar \cite{pions}.

Valid conclusions about the applicability of our approach can
obviously only be drawn in comparison to experimental data.  
In view of present experimental attempts to measure the coherent 
photoproduction of $\eta$ mesons on $^{4}{\rm He}$,
we also performed calculations for this nucleus, despite of the
limited applicability of our model assumptions for such a light
system. We have found that the cross section 
on $^{4}{\rm He}$
is dominated by the local
$\omega$-graph.

\acknowledgments 
We gratefully acknowledge helpful discussions with
M.~Benmerrouche and L.J.~Abu-Raddad.  We also would like to thank
Thomas Feuster for performing the elementary cross section
calculations.

\begin{table}
%\begin{tabular}{r|r|l|r|c|l|r|c|l|} 
\bea
\begin{array}{|rcl|crcl|crcl|} 
\hline \hline
{g_{_{\eta NN}}}            &=& 2.24 & \quad&
g_{_{\eta NS_{11}}}         &=& 2.1  & \quad& 
g_{_{\eta ND_{13}}}         &=& 6.76             \\
g_{_{\omega \eta \gamma}}   &=& 0.31 & \quad& 
g_{_{\gamma pS_{11}}}       &=& 0.73 & \quad& 
g_{_{\gamma pD_{13}}}^{(1)} &=& 5.46           \\
g_{\omega NN}^v             &=& 10   & \quad& 
g_{_{\gamma nS_{11}}}       &=&-0.62 & \quad& 
g_{_{\gamma nD_{13}}}^{(1)} &=&-0.97           \\
g_{\omega NN}^t             &=& 1.59 & \quad& 
                            & &      & \quad& 
g_{_{\gamma pD_{13}}}^{(2)} &=& 5.76           \\
                            & &      & \quad& 
                            & &      & \quad& 
g_{_{\gamma nD_{13}}}^{(2)} &=& 0.66           \\
\hline \hline 
\end{array}\nn
\eea
%\end{tabular} 
\caption { {Coupling constants used in this study.}}
\label{cc}
\end{table}

\begin{table}
%\centerline{
\begin{tabular}{|c|c|c|c|c|c|c|} 
 Nucleus & $V _v$ & $ r_v $ & $ a_v $ & $V_s$ & $ r_s $ &
$a_s$  \\
     &(\footnotesize{MeV}) &
 (\footnotesize{fm}) & (\footnotesize{fm} ) & (\footnotesize{MeV}) &
  (\footnotesize{fm}) & (\footnotesize{fm}) \\ \hline
$^{12}{\rm C} $  & 385.7 & 1.056 & 0.427 & -470.4 & 1.056 & 0.447 \\ \hline
$^{40}{\rm Ca} $ & 348.1 & 1.149 & 0.476 & -424.5 & 1.149 & 0.506 \\ \hline
$^{4}{\rm He}  $ & 375.7 & 1.2   & 0.287 & -499.4 & 1.2   & 0.287 \\ 
\end{tabular} %}
\caption { 
Strengths, reduced radii and diffusivities for the relativistic scalar 
and vector mean-field potentials, respectively.
}
\label{nucpotp}
\end{table}

\renewcommand{\baselinestretch}{1.0}

\begin{references}
\bibitem{feuster} Th. Feuster and U. Mosel, \Journal{\NP}{A612}{375}{1997}.

\bibitem{mukop} M. Benmerrouche, N.C. Mukhopadhyay and J.F. Zhang,
\Journal{\PR}{D51}{3237}{1995}.

\bibitem{trya} V.A. Tryasuchev and A.I. Fiks, 
{\it Phys. Atom. Nucl.} {\bf 58} (1995) 1168.
  
\bibitem{bennhold} C. Bennhold  and H. Tanabe, \Journal{\NP}{A530}{625}{1991}, 
  \Journal{\PL}{B243}{13}{1990}
  
\bibitem{tiator} L. Tiator, C. Bennhold, S.S. Kamalov,
  \Journal{\NP}{A580}{455}{1994}.

\bibitem{fix} A. Fix and H. Arenh\"ovel, \Journal{\NP}{A620}{457}{1997}.

\bibitem{pieka} J. Piekarewicz, A.J. Sarty and M. Benmerrouche, 
  \Journal{\PR}{C55}{2571}{1997},
L.J. Abu-Raddad, J. Piekarewicz, A.J. Sarty 
and M. Benmerrouche, \Journal{\PR}{C57}{2053}{1998}.


\bibitem{pions} W. Peters, H. Lenske and U. Mosel, nucl-th 9803009, submitted to
  Nucl. Phys. A.

\bibitem{bernd} B. Krusche et al., \Journal{\PRL}{74}{3736}{1995}.

\bibitem{who} F. de Jong und H. Lenske, 
\Journal{\PR}{C54}{1488}{1996};
H. M\"uther, G. Knehr und A. Polls, 
  \Journal{\PR}{C52}{2955}{1995}.

\bibitem{joach} C. J. Joachain, {\it Quantum collision theory}, 3rd
  edition, North Holland Physics Publishing, Amsterdam 1983.

\bibitem{carrasco} R.C. Carrasco, 
  \Journal{\PR}{C48}{2333}{1993}.

\bibitem{johan} J.I. Johansson and H.S. Sherif, 
  \Journal{\NP}{A575}{477}{1994}.

\bibitem{li} X. Li , L.E. Wright and C. Bennhold, 
  \Journal{\PR}{C48}{816}{1993}.

\bibitem{tiator2} L. Tiator and L.E. Wright, 
  \Journal{\PR}{C30}{989}{1984}.

\bibitem{chiang} H.C. Chiang, E. Oset and L.C. Liu, 
 \Journal{\PR}{C44}{738}{1991}.

\bibitem{effe} M. Effenberger and A. Sibirtsev, 
 \Journal{\NP}{A632}{99}{1998}.

\bibitem{negele} J.W. Negele and D. Vautherin, \Journal{\PR}{C5}{1472}{1972}.

\bibitem{bernd2} B. Krusche , {\it private communication.}

\bibitem{schia} R. Schiavella, V.R. Pandharipande and R.B. Wiringa,
 \Journal{\NP}{A449}{219}{1986}.

\end{references}
\end{document}